\documentclass[10pt]{article}
\usepackage{amsmath,amssymb}

\newcommand{\be}{\begin{equation}}
\newcommand{\ee}{\end{equation}}
\newcommand{\bea}{\begin{eqnarray}}
\newcommand{\eea}{\end{eqnarray}}

\newcommand{\lbl}[1]{\label{eq:#1}}
\newcommand{ \rf}[1]{(\ref{eq:#1})}

\newskip\humongous \humongous=0pt plus 1000pt minus 1000pt

\newif\ifdtup

\newcommand{\lapprox}{%
\mathrel{%
\setbox0=\hbox{$<$}
\raise0.6ex\copy0\kern-\wd0
\lower0.65ex\hbox{$\sim$}
}}
\newcommand{\gapprox}{%
\mathrel{%
\setbox0=\hbox{$>$}
\raise0.6ex\copy0\kern-\wd0
\lower0.65ex\hbox{$\sim$}
}}

\def\theequation{\arabic{section}.\arabic{equation}}

\begin{document}

\begin{center}

{\bf\Large{On some properties of the fourth-rank \\
hadronic vacuum polarization tensor\\[0.15cm]
and the anomalous magnetic moment of the muon }}\\[0.6cm]

{Marc Knecht}

\vspace*{0.4cm}
Centre~de~Physique~Th\'{e}orique,\\ 
CNRS/Aix-Marseille Univ./Univ.~de~Toulon (UMR 7332)\\
CNRS-Luminy Case 907, 13288 Marseille Cedex 9, France\\
\vspace*{0.5cm}

\begin{abstract}
  Some short-distance properties of the fourth-rank  
  hadronic vacuum polarization tensor are re-examined.
  Their consequences are critically discussed in the context of
  the hadronic light-by-light scattering contribution
  to the anomalous magnetic moment of the muon.
\end{abstract}


\end{center}



\section{Introduction}
\label{sec:intro}
\setcounter{equation}{0}

The Muon g-2 Collaboration is about to release, 
some 15 years after the final publication \cite{Bennett:2006fi} of the BNL-E821 experiment, 
the first result on a new high-precision measurement,  conducted by 
the FNAL-E989 experiment, of the anomalous magnetic moment of the muon
$a_\mu$. It is thus not surprising that quite some theoretical activity 
aiming at improving the standard model prediction for this observable 
is going on. The main limitations on this endeavour come from the hadronic 
contributions, hadronic vacuum polarization (HVP) and hadronic light-by-light 
scattering (HLxL). The former is traditionally evaluated through a dispersion 
relation, whose absorptive part is determined directly from data on $e^+ e^- \to {\rm hadrons}$.
The most recent evaluations \cite{Jegerlehner:2017lbd,Davier:2019can,Keshavarzi:2019abf}
along these lines have now reached a
precision that, in relative terms, lies below the $0.5\%$ level. The determination
of HVP from numerical simulations of QCD on a lattice has been developing
fast in recent years. Recent results \cite{DellaMorte:2017dyu,Blum:2018mom,Giusti:2019xct,Davies:2019efs,Gerardin:2019rua,Borsanyi:2020mff}, 
although they have not yet reached the same level of precision than the traditional approach, look promising.
Finally, the MUonE proposal \cite{MUonE_LoI}, which aims at an experimental evaluation
of HVP directly in the space-like region, and in an inclusive manner,
could be an interesting complementary alternative for the future, although both
theoretical and experimental challenges are high \cite{Calame:2015fva,Abbiendi:2016xup,Banerjee:2020tdt}.

Assuming that the results for the HVP contribution obtained through
these various approaches will eventually agree and reach comparable 
precisions, HLxL will then stand out as the main source of theoretical uncertainty,
hence the many recent efforts devoted to its evaluation. 
Here also, various approaches are being considered and developed, ranging from
lattice simulations of QCD \cite{Blum:2015gfa,Blum:2015you,Blum:2016lnc,Blum:2019ugy,Gerardin:2016cqj,Gerardin:2019vio}
to dispersion relations \cite{Pauk:2014rfa,Colangelo:2015ama,Colangelo:2017fiz}, 
by way of, to mention but a few, five-dimensional models \cite{Leutgeb:2019zpq,Leutgeb:2019gbz,Cappiello:2019hwh}, 
Schwinger-Dyson equations \cite{Eichmann:2019tjk,Raya:2019dnh}, the Schwinger sum rule \cite{Hagelstein:2017obr,Hagelstein:2019tvp}
and various dispersive or phenomenological approaches, often devoted to estimating a specific contribution 
(e.g. various single-meson poles) only
\cite{Knecht:2001qf,Nyffeler:2009tw,Pauk:2014rta,Roig:2014uja,Masjuan:2017tvw,Guevara:2018rhj,Hoferichter:2018dmo,Knecht:2018sci,Hoferichter:2018kwz,%
Roig:2019reh,Colangelo:2019lpu,Colangelo:2019uex,Melnikov:2019xkq}. 
For recent surveys and more extended lists of references, see Refs. \cite{Jegerlehner:2017gek,Meyer:2018til,Danilkin:2019mhd}.
Reducing the theoretical relative uncertainty of the HLxL contribution to a {\it reliable} 
level of $\sim 10 \%$ would already constitute a remarkable achievement.

This note is devoted to some aspects related to one of the specific
contributions mentioned above, namely the one due to the pion pole.
There are essentially two reasons that explain why this particular contribution to HLxL 
has attracted so much attention in the past, and keeps on being a point of focus even today. 
First, in the limit where the number of colours $N_c$ becomes large \cite{tHooft:1973alw}, 
only single-meson exchanges are relevant \cite{Witten:1979kh}, and the pion being the lightest meson, it is
expected to provide the main contribution \cite{deRafael:1993za,Knecht:2001qg}. Second, in the first serious 
attempts \cite{Hayakawa:1995ps,Bijnens:1995cc,Bijnens:1995xf,Hayakawa:1996ki,Hayakawa:1997rq,Hayakawa:2001bb,Bijnens:2001cq}
to perform a complete evaluation of the HLxL contribution to $a_\mu$, it so occurred
that the final result was in fact almost completely given by the 
contribution due to the pion pole, the other contributions cancelling
almost exactly among thenselves. This cancellation took place
although the hadronic models considered in the various studies were 
actually exhibiting quite different features, as discussed, for instance, 
in Ref. \cite{Prades:2001zv}. Thus, having the pion-pole 
contribution under good control is currently considered to be an essential step 
into the direction of obtaining an accurate and reliable evaluation of the HLxL component of $a_\mu$.

In the past various authors 
have used different definitions of what they have chosen to call the
contribution from the ``pion pole'', see the discussion in the review \cite{Jegerlehner:2009ry}. 
At present, this issue does no longer seem to constitute a point of contention.
Recently a different debate concerning the pion-pole 
contribution to HLxL has resurfaced in the literature \cite{Colangelo:2019lpu,Colangelo:2019uex,Melnikov:2019xkq}. 
It has been triggered mainly because of different views as how to implement a certain short-distance
constraint, first obtained in Ref. \cite{Melnikov:2003xd}, on the rank-four hadronic vacuum polarization tensor, which is the 
central object for HLxL. These different views then lead to quite
different numerical evaluations of the pion-pole contribution, see for instance the discussion in Ref. \cite{Colangelo:2019uex}. 
Although the pion pole is only one contribution to HLxL among many, 
and what should actually matter in the end is the full contribution to $a_\mu$ from HLxL,
it is certainly of interest, given the importance of this contribution, 
to understand what are the whys and wherefores of this rather confusing situation.

Thus, the purpose of this note is therefore not to provide 
yet another new evaluation of the HLxL contribution. Rather, it was written
with the aim of scrutinizing this particular issue in greater detail and, possibly, of
providing some understanding that may contribute to settle it.
The outline of the remaining part of the text is as follows. First, I recall,
in Section \ref{sec:properties}, general properties of the four- and three-point functions 
relevant for this discussion, including the short-distance
condition that relates them. I then describe in detail the implementation of this condition 
in Section \ref{sec:consequences} in general, before focusing on its
implications for the contribution of the pseudoscalar poles. Finally, I give
a summary and conclusions in Section \ref{sec:summary}. Some more technical aspects
related to the short-distance expansion have been gathered in Appendix A for the interested reader. 
Appendix B illustrates the discussion from the perspective of the 
low-energy expansion.

\section{Some hadronic four- and three-point functions and their properties}
\label{sec:properties}
\setcounter{equation}{0}

As mentioned in the introduction, the central object of interest is the connected four-point QCD correlator
\be
{\cal W}_{\mu\nu\rho\sigma}(q_1,q_2,q_3, q_4) =
\frac{1}{i} \int \!\! d^4x_1 \! \int \!\! d^4x_2 \! \int \!\! d^4x_3
e^{i(q_1\cdot x_1 + q_2\cdot x_2 + q_3\cdot x_3)}
\langle\Omega\vert T \! \left\{
j_\mu (x_1) j_\nu (x_2) j_\rho (x_3) j_\sigma (0) \right\}_{\!C}
\vert \Omega \rangle
,
\ee
where $j_\mu (x)$ stands for the light-quark component of the hadronic part of the electromagnetic current,
\be
j_\mu = {\bar\psi} Q \gamma_\mu \psi  ,
\, \psi = \left(
\begin{tabular}{c}
$ u $ \\
$ d $ \\
$ s $
\end{tabular}
\right)
,
\, 
Q = {\rm diag} \left( + \frac{2}{3} , - \frac{1}{3} , - \frac{1}{3} \right)
,
\lbl{current}
\ee
and $\vert \Omega \rangle$ denotes the QCD vacuum. For notational convenience, I have
written this correlator as a function of four variables, but only three momenta are
actually independent, since invariance under tranlations requires that the condition
\be
q_1 + q_2 + q_3 + q_4 = 0
\lbl{mom_sum}
\ee
holds.
Let me recall that $a_\mu^{\mbox{\tiny{HLxL}}}$, the HLxL contribution to the anomalous 
magnetic moment of the muon, can be expressed in terms of this correlator in the following 
way \cite{Aldins:1970id}
\be
a_\mu^{\mbox{\tiny{HLxL}}} \, \equiv \, 
\frac{1}{48 m_\ell}\,{\mbox{tr}}[ (\not\! p + m_{\ell}) [ \gamma^\sigma ,\gamma^\tau ]
(\not\!p+m_{\ell}) \Gamma_{\sigma\tau}^{\mbox{\tiny{HLxL}}}(p,p) ]   ,
\lbl{a_mu_HLxL_1}
\ee
where $p$ stands for the momentum of the muon and $\Gamma_{\sigma\tau}^{\mbox{\tiny{HLxL}}}(p,p)$
is the limit of the vertex function $\Gamma_{\sigma\tau}^{\mbox{\tiny{HLxL}}}(p',p)$, defined as
\bea
\overline{\mbox{u}}(p')\Gamma_{\sigma\tau}^{\mbox{\tiny{HLxL}}}(p',p)\mbox{u}(p) 
&=&
e^6\,\int\frac{d^4q_1}{(2\pi)^4}\int\frac{d^4q_2}{(2\pi)^4}\,,
\frac{1}{q_1^2\,q_2^2\,(q_1+q_2-k)^2}\,
\nonumber\\
&&\quad
\times
\frac{1}{(p'-q_1)^2-m_{\ell}^2}\,
\frac{1}{(p'-q_1-q_2)^2-m_{\ell}^2}
\nonumber\\
&&\quad
\times
\overline{\mbox{u}}(p')
\gamma^{\mu}(\not\! p'- \not\!q_1+m_{\ell})
\gamma^{\nu}(\not\! p'- \not\! q_1-\not\! q_2+m_{\ell})
\gamma^{\rho}\mbox{u}(p)
\nonumber\\
&&\quad
\times
{\cal W}_{\mu\nu\rho\sigma\tau}(q_1,q_2,k-q_1-q_2,-k)  ,
\lbl{a_mu_HLxL_2}
\eea
when the momentum difference $k=p'-p$ vanishes. This definition involves the derivative 
of the four-point function,
\be\lbl{WI1}
{\cal W}_{\mu\nu\rho\sigma\tau}(q_1,q_2,k - q_1 - q_2, -k) \equiv
\frac{\partial}{\partial k^\sigma} {\cal W}_{\mu\nu\rho\tau}(q_1,q_2,k-q_1-q_2,-k) ,
\ee
with respect to its fourth momentum $k$. Eq. \rf{a_mu_HLxL_1} then requires to take the limit $k\to 0$
of this derivative. Due to the conservation of the current $j_\mu$,
the rank-four hadronic vacuum polarization tensor satisfies the Ward identities
\be
\{ q_{1\mu} ; q_{2\nu} ; q_{3\rho} ; q_{4\sigma} \} {\cal W}^{\mu\nu\rho\sigma}(q_1,q_2,q_3,q_4)
= \{ 0 ; 0 ; 0 ; 0\}  .
\ee
Based on these transversality properties combined with Bose symmetry,
the authors of Ref. \cite{Colangelo:2015ama} have obtained a decomposition of the tensor
${\cal W}^{\mu\nu\rho\tau}$,
\be
{\cal W}^{\mu\nu\rho\tau}(q_1,q_2,q_3,q_4) = \sum_{i=1}^{54} {\cal W}_i (q_1,q_2,q_3,q_4)
T_i^{\mu\nu\rho\tau} (q_1,q_2,q_3,q_4).
\lbl{W4_decomp}
\ee
in terms of invariant functions ${\cal W}_i (q_1,q_2,q_3,q_4)$ free from
kinematic singularities and zeroes. These functions actually depend on the 
invariants that can be built with the products $q_i \cdot q_j$, $i,j=1,2,3,4$,
but for simplicity I write them as functions of the momenta for the time being.
Not much is known about these functions beyond the kinematic properties
mentioned above, and in order to estimate them, or at least the subset
of them that contributes to $a_\mu^{\mbox{\tiny{HLxL}}}$, it is important
to make sure that they satisfy the few properties that can be deduced
directly from QCD. One of these properties arises from the well-known
behaviour \cite{Bjorken:1966jh,Johnson:1966se} of the time-ordered product 
of two currents \rf{current} at short distances,
\be
\lim_{q\to\infty} \int d^4 x \, e^{i q \cdot x} T \{ j_\mu (x) j_\nu (0) \} =
-2 \epsilon_{\mu\nu\alpha\beta} \frac{q^\alpha}{q^2} A^\beta (0) + {\cal O} (q^{-2})   ,
\lbl{BJL_limit}
\ee
where it is understood here and in what follows that the limit holds
when the momentum $q$ belongs to the Euclidian region and when all 
its components become simultaneously large. The axial current appearing
on the right-hand side of this relation is defined as
\be
A_\mu \equiv {\bar\psi} Q^2 \gamma_\mu \gamma_5 \psi = \sum_{a=3,8,0} \!\! {\rm tr}(Q^2\lambda^a) A_\mu^a  ,
\ A_\mu^a \equiv {\bar\psi} \frac{\lambda^a}{2} \gamma_\mu \gamma_5 \psi  ,
\ \frac{\lambda^0}{2} \equiv  \frac{{1\!\!1}}{\sqrt{6}}   .
\lbl{axial_currents}
\ee
Then, writing
\be
q_1 = {\bar q} + {\hat q} , \ q_2 = {\bar q} - {\hat q} , \ {\hat q}^2 = - Q^2 , \ Q^2 >0  ,
\ee
one establishes \cite{Melnikov:2003xd} the following short-distance behaviour when the momenta carried by the first 
two currents become hard, while the other two remain soft (note that $q_3+q_4=-q_1-q_2$ remains soft as well),
\be
{\cal W}_{\mu\nu\rho\sigma}({\bar q} + {\hat q},{\bar q} - {\hat q},q_3,q_4) = 
-  2 \, \epsilon_{\mu\nu\tau\alpha} \frac{{\hat q}^\alpha}{{\hat q}^2} \, {\cal W}_{\rho\sigma}^{\ \ \, \tau} (q_3 , q_4)
+ {\cal O} ( {\hat q}^{-2} )  ,
\lbl{MV_SD}
\ee
where the three-point function ${\cal W}_{\mu\nu\rho} (q_1 , q_2)$ is defined as
\be
{\cal W}_{\mu\nu\rho} (q_1 , q_2) = i 
\int \!\! d^4x_1 \! \int \!\! d^4x_2 \,
e^{i(q_1\cdot x_1 + q_2\cdot x_2)}
\langle\Omega\vert T \! \left\{
j_\mu (x_1) j_\nu (x_2) A_\rho (0) \right\}
\vert \Omega\rangle  .
\lbl{def_VVA}
\ee
It  satisfies the Ward identities
\be
\{ q_1^{\mu} ; q_2^{\nu} \} {\cal W}_{\mu\nu\rho} (q_1 , q_2)
= \{ 0 ; 0 \}  ,
\ ( q_{1} + q_{2} )^\rho {\cal W}_{\mu\nu\rho} (q_1 , q_2)
 =
 \mathcal{A} \, \epsilon_{\mu\nu\alpha\beta} q_1^{\alpha} q_2^{\beta}
 + {\cal W}_{\mu\nu} (q_1 , q_2)
 ,
 \lbl{WI_anom}
\ee
where ${\cal A}$ stands for the anomalous contribution \cite{Adler:1969gk,Bell:1969ts}
\be
{\cal A} = - \frac{N_c}{2\pi^2} \, {\rm tr} \, Q^4 = - \frac{N_c}{9 \pi^2}  .
\ee 
These Ward identities feature yet another three-point function,
\be
{\cal W}_{\mu\nu} (q_1 , q_2) \equiv
\int \!\! d^4 x_1 \int \!\! d^4 x_2 \, e^{i\left( q_1 \cdot x_1 + q_2 \cdot x_2 \right)}
\langle {\rm vac} \vert T \{ 
j_{\mu} (x_1) j_{\nu} (x_2) [D (0) + \frac{\alpha_s}{6 \pi} (G \cdot {\widetilde G}) (0) ]
\} \vert {\rm vac} \rangle
,
\ee
with
\be
D \equiv {\bar\psi} \{Q^2 , {\cal M} \} i \gamma_5 \psi  ,
\ {\cal M} = {\rm diag} (m_u , m_d , m_s)  ,
\ (G \cdot {\widetilde G}) \equiv \frac{1}{2} \epsilon_{\mu\nu\rho\sigma} G^{\mu\nu} G^{\rho\sigma}  ,
\ee
where $m_q$, $q=u,d,s,$ denotes the masses of the three lightest quarks
and $G^{\mu\nu}$ is the gluon field strength.
The decomposition of ${\cal W}_{\mu\nu} (q_1 ; q_2)$ is quite simple, since it 
involves a single function that is also free of kinematic singularities,
\be
{\cal W}_{\mu\nu} (q_1 , q_2) = 
{\cal H} (q_1^2,q_2^2,(q_1+q_2)^2)
\, \epsilon_{\mu\nu\alpha\beta} q_1^\alpha q_2^\beta
.
\lbl{W_long_decomp}              
\ee
This representation is entirely fixed by Lorents covariance, Bose symmetry, invariance
under parity and conservation of the current $j_\mu$, which imposes
transversality,
\be
\{ q_1^{\mu} ; q_2^{\nu} \} {\cal W}_{\mu\nu} (q_1 , q_2) = \{ 0 ; 0 \}   .
\ee
Achieving a similar decomposition for the three-point function
${\cal W}_{\mu\nu\rho} (q_1 , q_2)$ is not quite as straightforward.
Using only Lorentz covariance, invariance under parity, Bose symmetry
and Schouten's identity to eliminate two additional possible structures,
$q_1^\nu \epsilon^{\mu\rho\alpha\beta} q_{1\alpha} q_{2\beta}
+ q_2^\mu \epsilon^{\nu\rho\alpha\beta} q_{1\alpha} q_{2\beta}$ and
$q_2^\nu \epsilon^{\mu\rho\alpha\beta} q_{1\alpha} q_{2\beta}
- q_1^\mu \epsilon^{\nu\rho\alpha\beta} q_{1\alpha} q_{2\beta}$,
one obtains, to start with, the general decomposition
\bea
{{\cal W}}_{\mu\nu\rho} ( q_1 , q_2)
\!\!&=&\!\!
\epsilon_{\mu\nu\alpha\beta} q^\alpha_1 q_2^\beta (q_1 + q_2 )_\rho W_0 (q_1^2 , q_2^2 , (q_1 + q_2)^2)
+
\epsilon_{\mu\nu\alpha\beta} q_1^\alpha q_2^\beta (q_1 - q_2 )_\rho W_1 (q_1^2 , q_2^2 , (q_1 + q_2)^2)
\nonumber\\
&&
\!\!\!\!\!\!\!\!
+\,
\left[ q_{1 \nu} \epsilon_{\mu\rho\alpha\beta} q_1^\alpha q_2^\beta
- q_{2 \mu} \epsilon_{\nu\rho\alpha\beta} q_1^\alpha q_2^\beta \right] W_2 (q_1^2 , q_2^2 , (q_1 + q_2)^2)
\nonumber\\
&&
\!\!\!\!\!\!\!\!
+\,
\left[ q_{1 \mu} \epsilon_{\nu\rho\alpha\beta} q_1^\alpha q_2^\beta
+ q_{2 \nu} \epsilon_{\mu\rho\alpha\beta} q_1^\alpha q_2^\beta \right] W_3 (q_1^2 , q_2^2 , (q_1 + q_2)^2)
\nonumber\\
&&
\!\!\!\!\!\!\!\!
+ \,
\epsilon_{\mu\nu\rho\alpha} (q_1 - q_2)^\alpha W_4 (q_1^2 , q_2^2 , (q_1 + q_2)^2)
+
\epsilon_{\mu\nu\rho\alpha} (q_1 + q_2)^\alpha W_5 (q_1^2 , q_2^2 , (q_1 + q_2)^2)
\qquad~
\lbl{W_i_def}
\eea
in terms of six amplitudes that are free of kinematic singularities.
The use of Schouten's identity, as well as Bose symmetry, may well introduce
kinematic zeroes, but this issue is not relevant for our present purposes,
so I will not take it into consideration. Bose symmetry further requires
\be
W_i (q_2^2 , q_1^2 , (q_1 + q_2)^2) = (-1)^i W_i (q_1^2 , q_2^2 , (q_1 + q_2)^2)
,
\ i=0,1,2,3,4,5
.
\ee
Conservation of the electromagnetic current implies
\be
W_5 (q_1^2 , q_2^2 , (q_1 + q_2)^2) 
+ (q_1^2 + q_2^2)  W_3 (q_1^2 , q_2^2 , (q_1 + q_2)^2) = 0
,
\ee
and
\be
2 W_4 (q_1^2 , q_2^2 , (q_1 + q_2)^2) 
- (q_1^2 - q_2^2) W_3 (q_1^2 , q_2^2 , (q_1 + q_2)^2) + 2 ( q_1 \cdot q_2) W_2 (q_1^2 , q_2^2 , (q_1 + q_2)^2) = 0
.
\ee
These identities allow to eliminate $W_4 (q_1^2 , q_2^2 , (q_1 + q_2)^2)$ and
$W_5 (q_1^2 , q_2^2 , (q_1 + q_2)^2)$ in terms of the remaining functions without
introducing kinematic singularities. The result reads
\be
{{\cal W}}^{\mu\nu\rho} ( q_1 , q_2) = \sum_{i=0}^3
W_i (q_1^2 , q_2^2 , (q_1 + q_2)^2) \tau_{i}^{ \mu\nu\rho} (q_1 , q_2)
,
\lbl{calW_W_i}
\ee
with
\bea
\tau_0^{\mu\nu\rho} (q_1 , q_2 ) & = & \epsilon^{\mu\nu\alpha\beta} q_{1\alpha} q_{2\beta} (q_1 + q_2 )^\rho   ,
\nonumber\\
\tau_1^{\mu\nu\rho} (q_1 , q_2 ) & = & \epsilon^{\mu\nu\alpha\beta} q_{1\alpha} q_{2\beta} (q_1 - q_2 )^\rho   ,
\nonumber\\
\tau_2^{\mu\nu\rho} (q_1 , q_2 ) & = & q_1^\nu \epsilon^{\mu\rho\alpha\beta} q_{1\alpha} q_{2\beta}
- q_2^\mu \epsilon^{\nu\rho\alpha\beta} q_{1\alpha} q_{2\beta} 
- (q_1 \cdot q_2) \epsilon^{\mu\nu\rho\alpha} (q_1 - q_2)_\alpha   ,
\nonumber\\
\tau_3^{\mu\nu\rho} (q_1 , q_2 ) & = & q_1^\mu \epsilon^{\nu\rho\alpha\beta} q_{1\alpha} q_{2\beta}
+ q_2^\nu \epsilon^{\mu\rho\alpha\beta} q_{1\alpha} q_{2\beta}
- q_1^2 \epsilon^{\mu\nu\rho\alpha} q_{2\alpha}
- q_2^2 \epsilon^{\mu\nu\rho\alpha} q_{1\alpha}
.
\eea
An alternative but equivalent decomposition in terms of four functions free of kinematical 
singularities can also be found in Eq. (4.9) of Ref. \cite{Jackiw85}.
The condition \rf{WI_anom} on $(q_1 + q_2)^\rho {\cal W}^a_{\mu\nu\rho} ( q_1 ; q_2)$ further requires
\bea
\lefteqn{
(q_1 + q_2)^2 [W_0 (q_1^2 , q_2^2 , (q_1 + q_2)^2) + W_2 (q_1^2 , q_2^2 , (q_1 + q_2)^2)]
}
\hspace{8.0cm}~
\nonumber\\
\lefteqn{
+ \, (q_1^2 - q_2^2) \! \left[ W_1 (q_1^2 , q_2^2 , (q_1 + q_2)^2) - W_3 (q_1^2 , q_2^2 , (q_1 + q_2)^2) \right]
}
\hspace{8.3cm}~
\nonumber\\
\lefteqn{
- \, ( q_1^2 + q_2^2 ) W_2^a (q_1^2 , q_2^2 , (q_1 + q_2)^2) - {\mathcal A} - {\cal H} (q_1^2 , q_2^2 , (q_1 + q_2)^2)
= 0   ,
}
\hspace{8.3cm}~
\lbl{cond_W_i}
\eea
when combined with Eq. \rf{W_long_decomp}.
Expressing $W_0 (q_1^2 , q_2^2 , (q_1 + q_2)^2) $ in terms of the remaining functions through
this relation leads to the decomposition given in Ref. \cite{Knecht:2003xy}, with a slightly different notation, 
\be
{\cal W}_{\mu\nu\rho} (q_1 , q_2) = 
\frac{(q_1 + q_2 )_\rho}{(q_1 + q_2)^2}
\left[
{\mathcal A} + {\cal H} (q_1^2,q_2^2,(q_1+q_2)^2) \right] \epsilon_{\mu\nu\alpha\beta} q_1^{\alpha} q_2^{\beta}
+ \sum_{i=1}^3
w_i(q_1^2,q_2^2,(q_1+q_2)^2)\,t_i^{\mu\nu\rho}(q_1,q_2)
\lbl{W-decomp}
,
\ee 
in terms of a set of three fully transverse tensors $t_i^{\mu\nu\rho}(q_1 , q_2)$,
\bea
t_1^{\mu\nu\rho} (q_1 , q_2) & = & \tau_1^{\mu\nu\rho} (q_1 , q_2) -  \frac{q_1^2 - q_2^2}{(q_1 + q_2)^2} \tau_0^{\mu\nu\rho} (q_1 , q_2)   ,
\nonumber\\
t_2^{\mu\nu\rho} (q_1 , q_2) & = & \tau_2^{\mu\nu\rho} (q_1 , q_2) - \tau_0^{\mu\nu\rho} (q_1 , q_2) + \frac{q_1^2 + q_2^2}{(q_1 + q_2)^2} \tau_0^{\mu\nu\rho} (q_1 , q_2)  ,
\nonumber\\
t_3^{\mu\nu\rho} (q_1 , q_2) & = & \tau_1^{\mu\nu\rho} (q_1 , q_2) + \tau_3^{\mu\nu\rho} (q_1 , q_2)   ,
\eea
and with
\bea
w_1 (q_1^2 , q_2^2 , (q_1 + q_2)^2) & = & W_1 (q_1^2 , q_2^2 , (q_1 + q_2)^2) - W_3 (q_1^2 , q_2^2 , (q_1 + q_2)^2)  ,
\nonumber\\
w_2 (q_1^2 , q_2^2 , (q_1 + q_2)^2) & = & W_2 (q_1^2 , q_2^2 , (q_1 + q_2)^2)  ,
\nonumber\\
w_3 (q_1^2 , q_2^2 , (q_1 + q_2)^2) & = & W_3 (q_1^2 , q_2^2 , (q_1 + q_2)^2)   .
\lbl{w_vs_W}
\eea
But this elimination is done at the expense of introducing kinematic singularities into the tensors
$t_i^{\mu\nu\rho}(q_1 , q_2)$, and hence a kinematic constraint on the functions $w_i (q_1^2 , q_2^2 , (q_1 + q_2)^2)$.
Indeed, Eq. \rf{cond_W_i} precisely materializes this constraint, since it states that the combination
\be
(q_1^2 + q_2^2) w_2 (q_1^2 , q_2^2 , (q_1 + q_2)^2) - 
(q_1^2 - q_2^2) w_1 (q_1^2 , q_2^2 , (q_1 + q_2)^2)
+ {\mathcal A} + {\cal H} (q_1^2 , q_2^2 , (q_1 + q_2)^2)
\ee
has to be equal to $(q_1 + q_2)^2$ times a function free of any kinematic singularity.

Since the authors of Ref. \cite{Colangelo:2019uex} use the notation of Ref. \cite{Knecht:2003xy},
let me, before closing this section, provide the connection between the two.
It is straightforward to establish the relations
\be
t^{(+)\mu\nu\rho} (q_1 , q_2) = t_2^{\mu\nu\rho} (q_1 , q_2)  ,
\quad  
t^{(-)\mu\nu\rho} (q_1 , q_2) = t_1^{\mu\nu\rho} (q_1 , q_2)  ,
\quad  
{\tilde t}^{(-)\mu\nu\rho} (q_1 , q_2) = t_3^{\mu\nu\rho} (q_1 , q_2)  ,
\ee
Then, upon writing \cite{Knecht:2003xy}
\bea
{{\cal W}}_{\mu\nu\rho} ( q_1 , q_2) \!\!&=&\!\!
- \frac{1}{8 \pi^2}
\Big[ - w_L (q_1 , q_2) \tau_{0\mu\nu\rho} (q_1 , q_2) + w_T^{(+)} (q_1 , q_2) t^{(+)}_{\mu\nu\rho} (q_1 , q_2)
\nonumber\\
&& \qquad
+ \, w_T^{(-)} (q_1 , q_2) t^{(-)}_{\mu\nu\rho} (q_1 , q_2) + {\tilde w}_T^{(-)} (q_1 , q_2) {\tilde t}^{(-)}_{\mu\nu\rho} (q_1 , q_2)
\Big]   ,
\eea 
one obtains
\be 
8 \pi^2 w_1 (q_1 , q_2) = - w_T^{(-)} (q_1 , q_2)  ,
\quad 
8 \pi^2 w_2 (q_1 , q_2) = - w_T^{(+)} (q_1 , q_2)  ,
\quad
8 \pi^2 w_3 (q_1 , q_2) = - {\tilde w}_T^{(-)} (q_1 , q_2)  ,
\ee
and, making, for convenience, the change of notation $W_0 \longrightarrow w_0$,
\bea
\frac{(q_1 + q_2)^2}{8\pi^2} \, w_L (q_1 , q_2) \!\!&=&\!\! (q_1 + q_2)^2 \left[ w_0 (q_1 , q_2) + w_2 (q_1 , q_2) \right]
+  (q_1^2 - q_2^2) w_1 (q_1 , q_2) - (q_1^2 + q_2^2) w_2 (q_1 , q_2)
\nonumber\\
\!\!&=&\!\!   {\mathcal A} + {\cal H} (q_1^2 , q_2^2 , (q_1 + q_2)^2) 
.
\lbl{cond_w_i}
\eea
It is clear from this relation that the function $ w_L (q_1^2 , q_2^2 , (q_1 + q_2)^2)$
does in general exhibit kinematic singularities.
At this stage, let me formulate two remarks:
\begin{itemize}
 \item All the above properties still hold if instead of considering the correlators
 involving the current $A_\mu$, I had replaced the latter by one of its components
 $A_\mu^a$ defined in Eq. \rf{axial_currents}, with the proviso that each function
 like $W_i$ or $w_i$ is endowed with a corresponding superscript $a$, where $a=3,8,0$, 
 and that the anomaous contribution ${\cal A}$ is replaced by ${\cal A}^a \equiv {\cal A} \, {\rm tr} (Q^2\lambda^a/2)/{\rm tr}Q^4$.
 Following common practice, I will refer to these three cases $a=3,8,0$ as the iso-triplet, octet, 
 and singlet channels, respectively.
 
 \item In the limit where $q_2$ vanishes, or equivalently in the combined limit
 $q_2^2 \to 0$, $(q_1+q_2)^2 \to q_2^2$, the relation \rf{cond_w_i} becomes  
\be
\frac{1}{8\pi^2} \, w_L (q_1^2 , 0 , q_1^2) = w_0 (q_1^2 , 0 , q_1^2) + w_1 (q_1^2 , 0 , q_1^2) 
= \frac{1}{q_1^2} \left[ {\mathcal A} + {\cal H} (q_1^2 , 0 , q_1^2)  \right]  . 
\lbl{wL_q2_vanishes}
\ee
Two observations can be made from this relation. The first is that the combination
$ {\mathcal A} + {\cal H} (q_1^2 , 0 , q_1^2)$ vanishes as ${\cal O}(q_1^2)$, a statement in which,
when restricted to the iso-triplet channel and with the anomaly removed, one recognizes
the Sutherland-Veltman theorem \cite{Sutherland:1967vf,Veltman67}, see also Refs. \cite{Shore:1992pm} and \cite{Jackiw85}.
The second observation is more relevant for the subject of this note: in the chiral limit, or in the combined chiral and large-$N_c$ limit 
in the case of the singlet channel, ${\cal H} (q^2 , 0 , q^2)$ vanishes, and the relation \rf{wL_q2_vanishes} reduces to the usual expression 
$ w_L (q^2 , 0 , q^2)/8\pi^2 = {\mathcal A}/q^2$. Although the combination $w_0 (q^2 , 0 , q^2) + w_1 (q^2 , 0 , q^2)$
tends to the same expression in this limit, the way it arises, and the physical content it conveys, is completely different. 
I will come back to this issue and its consequences later on.
\end{itemize}

Finally, let me mention that the first equality in Eq. \rf{cond_w_i} also appears 
as Eq. (B13) of Ref. \cite{Roig:2019reh}, but its implications have not been discussed
by the authors.

\section{Implementing the short-distance constraints}
\label{sec:consequences}
\setcounter{equation}{0}

Coming back to the short-distance behaviour given in Eq. \rf{MV_SD}, it
may now be rewritten as
\be
{\cal W}^{\mu\nu\rho\sigma}({\bar q} + {\hat q},{\bar q} - {\hat q},q_3,q_4) = 
- \, \frac{2}{{\hat q}^2} \sum_{i=0}^3 w_i (q_3^2, q_4^2 , (q_3+q_4)^2 ) K_i^{\mu\nu\rho\sigma} ({\hat q} , q_3 , q_4)
+ {\cal O} ( {\hat q}^{-2} )  ,
\lbl{MV_SD_ter}
\ee
where
\bea\lbl{K-tensors}
K_i^{\mu\nu\rho\sigma} ({\hat q} , q_3 , q_4) \!\!&\equiv&\!\! \epsilon^{\mu\nu\tau\alpha} {\hat q}_\alpha \tau_{i~~\tau}^{\rho\sigma} (q_3 , q_4) ~~ i=0,1,2 ,
\nonumber\\[-0.15cm]
\\[-0.15cm]
K_3^{\mu\nu\rho\sigma} ({\hat q} , q_3 , q_4) \!\!&\equiv&\!\! \epsilon^{\mu\nu\tau\alpha} {\hat q}_\alpha 
[ \tau_{1~~\tau}^{\rho\sigma} (q_3 , q_4) + \tau_{3~~\tau}^{\rho\sigma} (q_3 , q_4) ] .
\nonumber
\eea
The task that needs to be done next is to work out the consequences
of the short-distance constraint \rf{MV_SD_ter} on the invariant 
functions ${\cal W}_i(q_1,q_2,q_3,q_4)$ that describe the fourth-rank
vacuum polarization tensor as shown in Eq. \rf{W4_decomp}. A procedure 
through which this can be achieved is described in Appendix A. 
Here I will merely discuss, through one example, some of the consequences 
that follow from the condition \rf{MV_SD_ter}.

The example I wish to consider involves, following Ref. \cite{Colangelo:2017fiz}, the combination
\be
{\hat{\cal W}}_1 (q_1 , q_2 , q_3 , q_4) \equiv {\cal W}_1 (q_1 , q_2 , q_3 , q_4) - (q_1 \cdot q_2) {\cal W}_{47} (q_1 , q_2 , q_3 , q_4)   .
\lbl{W1hat}
\ee
For this combination, the short-distance constraint \rf{MV_SD_ter} requires the condition
\be
{\hat{\cal W}}_1 ({\bar q} + {\hat q},{\bar q} - {\hat q} , q_3 , q_4)  =
- \frac{2}{{\hat q}^2}  \left[ w_0 (q_3^2, q_4^2 , (q_3+q_4)^2  ) + w_1 (q_3^2, q_4^2 , (q_3+q_4)^2  ) \right] + {\cal O}({\hat q}^{-4})
\lbl{SD_W1}
\ee
to hold. Before considering some specific aspects of this relation, a few
general statements may be useful:
\begin{itemize}
 \item 
This condition holds as it stands, i.e. for all values of the invariants $q_3^2$, $q_4^2$ and $(q_3+q_4)^2$.
 \item
Both sides are free from kinematic singularities. Since such singularities are absent 
on the left-hand side by construction, none should show up on the right-hand side, which
is the case.
 \item
Dynamical singularities in the variables $q_3^2$, $q_4^2$ and $(q_3+q_4)^2$, 
i.e. poles due to single-particle  exchanges or cuts due to multi-particle exchanges, 
have to match on both sides; those present in the functions $w_0$ and  $w_1$ must correspond 
to singularities also present in ${\hat{\cal W}}_1$ and that moreover survive in the limit under consideration; likewise,
singularities in ${\hat{\cal W}}_1$ that have no counterpart in $w_0$ or $w_1$ must fall into the subleading 
contributions to the short-distance expansion.
 \item
Since the momenta $q_3$ and $q_4$ are generic (i.e. non-exceptional in the
sense of Weinberg's theorem \cite{Weinberg:1959nj}), the chiral limit can be taken on both sides; the same
holds for the large-$N_c$ limit, or for the combination of both limits.
 \item
The limit where in addition $q_3^2$  becomes large in the Euclidian 
region can also be taken on both sides, as long as the condition $-{\hat q}^2 \gg -q_3^2$ 
remains satisfied.
\end{itemize}

\indent

\begin{center}
\textit{1.~~Pion pole}
\end{center}

Let us now consider the contribution coming from the exchange of a single 
neutral pion. It produces in ${\hat{\cal W}}_1$ a pole in the variable $(q_3+q_4)^2$,
\be
{\hat{\cal W}}_1^{(\pi^0)} (q_1 , q_2 , q_3 , q_4) = 
- \frac{{\cal F}_{\pi\gamma^*\gamma^*} (q_1^2 , q_2^2 ) \, {\cal F}_{\pi\gamma^*\gamma^*} (q_3^2 , q_4^2 )}{(q_3 + q_4)^2 - M_\pi^2}
\lbl{W1_pi0_pole}
\ee
involving the pion transition form factor ${\cal F}_{\pi\gamma^*\gamma^*}$ defined as
\be
i \int \!\! d^4 x \, e^{i q \cdot x}
\langle \Omega \vert T \! \left\{j_\mu (x/2) j_\nu (-x/2) \right\} \vert \pi^0 (p) \rangle
=
\epsilon_{\mu\nu\alpha\beta} q^\alpha p^\beta {\cal F}_{\pi\gamma^*\gamma^*} ((p/2+q)^2 , (p/2-q)^2 )  ,
\lbl{pi0_TFF}
\ee
and where Bose symmetry means that the form factor is unchanged upon replacing $q$ by $-q$.
Notice that the above definition implies that
\be
\lim_{q \to \pm p/2} {\cal F}_{\pi\gamma^*\gamma^*} ((p/2+q)^2 , (p/2-q)^2 ) = {\cal F}_{\pi\gamma^*\gamma^*} (M_\pi^2 , 0) = {\cal F}_{\pi\gamma^*\gamma^*} (0 , M_\pi^2)  .
\ee
It differs of course from 
\be
\lim_{(p/2\pm q)^2 \to 0} {\cal F}_{\pi\gamma^*\gamma^*} ((p/2+q)^2 , (p/2-q)^2 ) = {\cal F}_{\pi\gamma^*\gamma^*} (M_\pi^2/2 + 2 q^2 ,0 )
 = {\cal F}_{\pi\gamma^*\gamma^*} (0,M_\pi^2/2 + 2 q^2).
\ee
A pole singularity similar to the one in ${\hat{\cal W}}_1$ also shows up on the right-hand side of Eq. \rf{SD_W1}, since
\be
w_0 (q_3^2 , q_4^2 , (q_3+q_4)^2) = {\rm tr}(Q^2 \lambda^3) \frac{F_\pi {\cal F}_{\pi\gamma^*\gamma^*} (q_3^2,q_4^2)}{(q_3 + q_4)^2 - M_\pi^2}
+ \cdots  ,
\lbl{w0_pi0_pole}
\ee
where the ellipsis stands for terms that are regular at $(q_3 + q_4)^2 = M_\pi^2$
and $F_\pi$ denotes the pion decay constants defined as
\be
\langle \Omega \vert A_\mu^3 (0) \vert \pi^0 (p) \rangle = i F_\pi p_\mu    .
\lbl{Fpi}
\ee
That the pion pole is located in the function $w_0$ and that it takes the form given above 
follows directly from the structure of the three-point function ${\cal W}^{\mu\nu\rho}$
as given in Eq. \rf{calW_W_i}, and from the structure of the two matrix elements in
Eqs. \rf{pi0_TFF} and \rf{Fpi}. The interested reader may actually check this property explicitly on the calculation 
of the functions $w_i$ at next-to-leading order in the low-emergy expansion presented in Appendix B.
According to the third item in the list that follows Eq. \rf{SD_W1}, this same pole singularity in $w_0$
has to be recovered in the asymptotic limit of ${\hat{\cal W}}_1 ({\bar q} + {\hat q},{\bar q} - {\hat q} , q_3 , q_4)$.
This requires
\be
\lim_{-{\hat q}^2 \to +\infty} {\cal F}_{\pi\gamma^*\gamma^*} ({\hat q}^2 , {\hat q}^2 ) = \frac{2}{3} \, \frac{F_\pi}{{\hat q}^2} + {\cal O} ({\hat q}^{-4})  ,
\lbl{OPE_TFF}
\ee
a property that is known to hold \cite{Novikov:1983jt,Nesterenko:1982dn}, and that also follows from the result given in Eq. \rf{BJL_limit}.
Furthermore, the compatibility, via the short-distance constraint \rf{SD_W1}, between the two expressions
\rf{W1_pi0_pole} and \rf{w0_pi0_pole} manifestly continues to hold in the chiral limit.

We may now consider the kinematic regime relevant for the evaluation of $a_\mu^{\mbox{\tiny{HLxL}}}$.
According to the formulas given in Eqs. \rf{a_mu_HLxL_1}, \rf{a_mu_HLxL_2}, and \rf{WI1}, this
involves taking the derivative of the rank-four vacuum polarization tensor with respect to $q_4$,
and then letting $q_4 \to 0$, taking the constraint \rf{mom_sum} into account. Since the tensors 
$T_i^{\mu\nu\rho\sigma} (q_1,q_2,q_3,q_4)$ are all at least linear in the momentum $q_4$, this limit 
can be rewritten as
\be
\lim_{q_4 \to 0}
\frac{\partial}{\partial q_4^\sigma} {\cal W}_{\mu\nu\rho\tau}(q_1,q_2,-q_4-q_1-q_2,q_4)
\equiv
\sum_i {\cal W}_i (q_1,q_2,q_3,q_4) \vert_{q_4=0} \times
\lim_{q_4 \to 0} \frac{\partial}{\partial q_4^\sigma} T_i^{\mu\nu\rho\tau}(q_1,q_2,-q_4-q_1-q_2,q_4)  .
\lbl{lim_usual}
\ee
As far as the short-distance constraint \rf{SD_W1} is concerned, this means that we 
need to compare the leading term in the short-distance expansion of
\bea
\lim_{q_4 \to 0} \, 
-  \frac{ {\cal F}_{\pi\gamma^*\gamma^*} (({\bar q} + {\hat q})^2 , ({\bar q} - {\hat q})^2 ) {\cal F}_{\pi\gamma^*\gamma^*} (q_3^2 , q_4^2 )}{(q_3 + q_4)^2 - M_\pi^2}
&=&
-  \frac{ {\cal F}_{\pi\gamma^*\gamma^*} (({\bar q} + {\hat q})^2 , ({\bar q} - {\hat q})^2 ) {\cal F}_{\pi\gamma^*\gamma^*} (M_\pi^2 , 0 )}{q_3^2 - M_\pi^2}
\nonumber\\
&=&
- \frac{2}{3} \, \frac{F_\pi}{{\hat q}^2} \, \frac{{\cal F}_{\pi\gamma^*\gamma^*} (M_\pi^2 , 0 )}{q_3^2 - M_\pi^2} + {\cal O}({\hat q}^{-4})
\eea
with
\be
\lim_{q_4 \to 0} \, - \frac{2}{{\hat q}^2} {\rm tr}(Q^2 \lambda^3) \frac{F_\pi {\cal F}_{\pi\gamma^*\gamma^*} (q_3^2,q_4^2)}{(q_3 + q_4)^2 - M_\pi^2}
=
- \frac{2}{{\hat q}^2} {\rm tr}(Q^2 \lambda^3) \frac{F_\pi {\cal F}_{\pi\gamma^*\gamma^*} (M_\pi^2,0)}{q_3^2 - M_\pi^2} .
\ee
The two expressions clearly match, and keep on doing so if one further takes the 
chiral limit, where one gains the additional information that 
${\cal F}_{\pi\gamma^*\gamma^*} (M_\pi^2,0) \to F_0 \stackrel{\rm o}{{\cal F}}_{\pi\gamma^*\gamma^*} (0,0) = 3{\cal A}/4$,
where
\be
\stackrel{\rm o}{{\cal F}}_{\pi\gamma^*\gamma^*} (q_3^2,q_4^2) = \lim_{m_q\to 0} {\cal F}_{\pi\gamma^*\gamma^*} (q_3^2,q_4^2)   ,
\ F_0 = \lim_{m_q\to 0} F_\pi    .
\ee

In the dispersive approach of Refs. \cite{Colangelo:2015ama,Colangelo:2017fiz,Colangelo:2019lpu,Colangelo:2019uex}, 
the invariant functions ${\cal W}_i(q_1,q_2,q_3,q_4)$ are first expressed in terms of a set of appropriate kinematic variables, 
namely
\be
s=(q_1+q_2)^2, \ t = (q_1+q_3)^2, \ q_1^2, \ q_2^2, \ q_3^2, \ q_4^2  .
\lbl{disp_var}
\ee
Accordingly, the functions describing the three-point function ${\cal W}_{\mu\nu\rho}$
are to be written as $w_i (q_3^2 , q_4^2 , s)$. This rewriting in terms of the variables 
\rf{disp_var} does not change the short-distance condition \rf{MV_SD_ter}, and the right-hand 
side involves the same sum, $w_0 (q_3^2, q_4^2 , s  ) + w_1 (q_3^2, q_4^2 , s)$, as before. 
The pion-pole contributions in Eqs. \rf{W1_pi0_pole} and \rf{w0_pi0_pole} also remain the same, 
up to the denominators that are now rewritten as $s-M_\pi^2$. As long as we give the different 
variables in Eq. \rf{disp_var} generic values, the whole discussion leading to the condition \rf{OPE_TFF} 
can be repeated again, mutatis mutandis. So let us therefore turn to the kinematic regime relevant 
for the evaluation of $a_\mu^{\mbox{\tiny{HLxL}}}$. Here the dispersive approach requires to consider 
the reduced kinematics defined in Ref. \cite{Colangelo:2015ama}, so that Eq. \rf{lim_usual} is replaced by
\be
\lim_{q_4 \to 0}
\frac{\partial}{\partial q_4^\sigma} {\cal W}^{\mu\nu\rho\tau}(q_1,q_2,-q_4-q_1-q_2,q_4)
\underset{\rm{Ref.\,[22]}}{\longrightarrow}
\sum_i {\cal W}_i (s,t,q_1^2,q_2^2,q_3^2,q_4^2) \Big\vert_{\genfrac{}{}{0pt}{}{s=q_3^2}{q_4^2=0}}^{\genfrac{}{}{0pt}{}{ }{t=q_2^2}} \times
\lim_{q_4 \to 0} \frac{\partial}{\partial q_4^\sigma} T_i^{\mu\nu\rho\tau}(q_1,q_2,-q_4-q_1-q_2,q_4) .
\lbl{lim_disp}
\ee
For the contribution from the pion pole to the left-hand side of Eq. \rf{SD_W1} we find (the pion-pole
contribution does not depend on the variable $t$)
\bea
\lim_{\genfrac{}{}{0pt}{}{s\to q_3^2}{q_4^2\to 0}} \, 
-  \frac{ {\cal F}_{\pi\gamma^*\gamma^*} (({\bar q} + {\hat q})^2 , ({\bar q} - {\hat q})^2 ) {\cal F}_{\pi\gamma^*\gamma^*} (q_3^2 , q_4^2 )}{s - M_\pi^2}
&=&
-  \frac{ {\cal F}_{\pi\gamma^*\gamma^*} (({\bar q} + {\hat q})^2 , ({\bar q} - {\hat q})^2 ) {\cal F}_{\pi\gamma^*\gamma^*} (q_3^2 , 0 )}{q_3^2 - M_\pi^2}
\nonumber\\
&=&
- \frac{2}{3} \, \frac{F_\pi}{{\hat q}^2} \, \frac{{\cal F}_{\pi\gamma^*\gamma^*} (q_3^2 , 0 )}{q_3^2 - M_\pi^2} + {\cal O}({\hat q}^{-4})   ,
\eea
whereas in the same limit its contribution to the right-hand side reads
\be
\lim_{\genfrac{}{}{0pt}{}{s\to q_3^2}{q_4^2\to 0}} \, - \frac{2}{{\hat q}^2} {\rm tr}(Q^2 \lambda^3) \frac{F_\pi {\cal F}_{\pi\gamma^*\gamma^*} (q_3^2,q_4^2)}{s - M_\pi^2} =
- \frac{2}{{\hat q}^2} {\rm tr}(Q^2 \lambda^3) \frac{F_\pi {\cal F}_{\pi\gamma^*\gamma^*} (q_3^2,0)}{q_3^2 - M_\pi^2}   .
\ee
The results for the two sides of the condition \rf{SD_W1} differ from the previous case,
since the second transition form factor now retains a dependence on $q_3^3$,
but what matters is that they perfectly match, and this matching persists in the chiral limit,
which can be taken without problem.

To summarize this discussion of the pion pole, I find that, in the chiral limit, the short-distance 
constraint \rf{SD_W1} leads to
\be
\lim_{q_4\to 0} \lim_{{\hat q}^2 \to - \infty} {\cal W}_1^{(\pi^0)}
= 
- \frac{2}{3} \,\frac{F_\pi}{{\hat q}^2} \, \frac{{\cal F}_{\pi\gamma^*\gamma^*} (0,0)}{q_3^2} + {\cal O}({\hat q}^{-4})
=  \frac{N_c}{18 \pi^2} \, \frac{1}{{\hat q}^2} \, \frac{1}{q_3^2} + {\cal O}({\hat q}^{-4})   .
\lbl{MV_limit}
\ee
in the case where the kinematic configuration corresponding to $q_4\to 0$, and considered by the authors
of Refs. \cite{Melnikov:2003xd,Melnikov:2019xkq}, is taken. In the kinematic configuration corresponding to the 
dispersive treatment of the pion pole advocated by the authors of Refs. \cite{Colangelo:2019lpu,Colangelo:2019uex},
it instead leads to
\bea
\lim_{\genfrac{}{}{0pt}{}{s\to q_3^2}{q_4^2\to 0}} \lim_{{\hat q}^2 \to - \infty} {\cal W}_1^{(\pi^0)}
=
- \frac{2}{3} \, \frac{F_\pi}{{\hat q}^2}  \frac{{\cal F}_{\pi\gamma^*\gamma^*} (q_3^2,0)}{q_3^2} + {\cal O}({\hat q}^{-4})
= - \frac{4}{3} \, \frac{F_\pi^2}{{\hat q}^2} \left( \frac{1}{q_3^2} \right)^2 \left[ 1 + {\cal O} (q_3^{-2}) \right] + {\cal O}({\hat q}^{-4})   . ~~~  
\lbl{disp-limit}
\eea
The second equality in this last equation holds when $q_3^2$ becomes large in 
the Euclidian region (but with $-{\hat q}^2\gg -q_3^2$), where the result \cite{Lepage:1979zb,Lepage:1980fj}
\be
\lim_{q_3^2\to -\infty} {\cal F}_{\pi\gamma^*\gamma^*} (q_3^2,0) = \frac{2 F_\pi}{q_3^2} + {\cal O}(q_3^{-4})
\ee
can be used.
Both limits are, as far as I can see, legitimate, in the sense that none reveals any incoherence. 
However, they will most likely lead to different 
numerical outcomes as far as the contribution of the pion pole to $a_\mu^{\mbox{\tiny{HLxL}}}$
is concerned. But this needs not be a problem per se since what matters in the end is the
comparison of the results obtained once all contributions to $a_\mu^{\mbox{\tiny{HLxL}}}$ have been added up.

\indent

\begin{center}
\textit{2.~~An apparent paradox and its solution}
\end{center}

The debate in the literature on $a_\mu^{\mbox{\tiny{HLxL}}}$ that has resurfaced recently \cite{Melnikov:2019xkq,Colangelo:2019uex}
takes its origin in the fact that Eq. \rf{SD_W1} is usually written in terms of the function $w_L$,
\be
{\hat{\cal W}}_1 (q_3^2,({\bar q} + {\hat q} + q_3)^2,({\bar q} + {\hat q})^2,({\bar q} - {\hat q})^2,q_3^2,0) = 
- \frac{2}{{\hat q}^2} \, \frac{1}{8 \pi^2} \, w_L (q_3^2, 0, q_3^2) + {\cal O}({\hat q}^{-4})
\quad~{\rm Refs.}~\cite{Melnikov:2019xkq,Colangelo:2019uex}   .
\lbl{SD_MV+CHHLS}
\ee
As discussed after Eq. \rf{cond_w_i}, this is quite legitimate in the limit 
appropriate for the discussion of $a_\mu^{\mbox{\tiny{HLxL}}}$, whether one 
considers it in the form \rf{lim_usual} or in the form \rf{lim_disp}. But
although $w_L (q_3^2, 0, q_3^2)/(8\pi^2)$ and $w_0 (q_3^2, 0, q_3^2) + w_1 (q_3^2, 0, q_3^2)$
are the same functions, they differ by their physical content, and this
difference lies at the heart of the debate. In order to explain this point, let 
me consider the chiral limit and consider the iso-triplet channel, see the first 
remark after Eq. \rf{cond_w_i} for the explanation of the nomenclature and the notation. 
The discussion in the octet channel is exactly the same,
with the $\eta$ meson playing the role of the pion, and extends to the singlet channel
and the $\eta'$ meson if in addition the large-$N_c$ limit is taken as well. 
In the chiral limit, the function $w_L^3$ is known exactly in QCD, for arbitrary kinematics,
\be
\lim_{m_q \to 0} \frac{1}{8\pi^2} \, w_L^3 (q_3^2 , q_4^2 , (q_3+q_4)^2) = \frac{3}{4} \, \frac{\cal A}{(q_3+q_4)^2}  ,
\lbl{w_L_chiral}
\ee
and this single contribution is entirely produced by a dynamical pion pole. Comparing the pion pole in ${\hat{\cal W}}_1$ 
with the one in $w_L^3/(8\pi^2)$ would lead to compare, in the ``dispersive'' limit \rf{lim_disp}
\be
- \frac{2}{3} \, \frac{F_0}{{\hat q}^2} \, \frac{\stackrel{\rm o}{{\cal F}}_{\pi\gamma^*\gamma^*} (q_3^2 , 0 )}{q_3^2} + {\cal O}({\hat q}^{-4}) 
~~{\rm vs.}~
- \frac{2}{3} \, \frac{F_0}{{\hat q}^2} \, \frac{\stackrel{\rm o}{{\cal F}}_{\pi\gamma^*\gamma^*} (0 , 0 )}{q_3^2} + {\cal O}({\hat q}^{-4})
=
- \frac{2}{3} \, \frac{1}{{\hat q}^2} \,  \frac{3}{4} \, \frac{\cal A}{q_3^2}   + {\cal O}({\hat q}^{-4})
.
\lbl{pseudo-paradox}
\ee
Clearly, the two expressions cannot match as such for all values of $q_3^2$,
and this mismatch is at the origin of the debate between the authors of Refs. \cite{Melnikov:2003xd,Melnikov:2019xkq}
on the one hand, and the authors of Refs. \cite{Colangelo:2019lpu,Colangelo:2019uex} on the 
other hand, the former seeing ``the dependence on this form factor 
[i.e. $\stackrel{\rm o}{{\cal F}}_{\pi\gamma^*\gamma^*} (q_3^2 , 0 )$] on $q_3^2$'' as 
``ambiguous within the dispersive approach'', whereas for the latter the model based 
on a constant form factor [i.e. $\stackrel{\rm o}{{\cal F}}_{\pi\gamma^*\gamma^*} (0 , 0 )$
in the chiral limit] represents a ``distorsion'' of the low-energy behaviour 
of the rank-four vacuum polarization tensor. But we have just seen that, although 
the two ways to implement the kinematic limit relevant for $a_\mu^{\mbox{\tiny{HLxL}}}$
give different results for the pion pole, they are both consistent with the content 
of Eq. \rf{SD_W1}, and the confrontation between the two options in Eq. \rf{pseudo-paradox}
never shows up.

In order to understand the origin of this apparent paradox, let us come back to the 
combination $w_0+w_1$ that actually appears on the right-hand side of Eq. \rf{SD_W1}.
Even in the chiral limit, the structure of this function remains quite different from the 
simple form taken by $w_L^3$ and given in Eq. \rf{w_L_chiral},
\bea\lbl{w_0+w_1_chiral}
\lim_{m_q\to 0} \left[ w_0^3 (q_3^2, q_4^2 , (q_3+q_4)^2  ) + w_1^3 (q_3^2, q_4^2 , (q_3+q_4)^2  ) \right]
\!\!&=&\!\!\!
\frac{F_0 \stackrel{\rm o}{{\cal F}}_{\pi\gamma^*\gamma^*} (q_3^2,q_4^2)}{(q_3 + q_4)^2} + \Delta w^3  (q_3^2, q_4^2 , (q_3+q_4)^2  )
\nonumber\\
\!\!&=&\!\!\! \frac{3{\cal A}/4}{(q_3+q_4)^2} 
+ \frac{F_0 [\stackrel{\rm o}{{\cal F}}_{\pi\gamma^*\gamma^*} (q_3^2,q_4^2) - \stackrel{\rm o}{{\cal F}}_{\pi\gamma^*\gamma^*} (0,0)]}{(q_3 + q_4)^2} 
\nonumber\\
&&\!\!\!\!\!\!
+ \, \Delta w^3  (q_3^2, q_4^2 , (q_3+q_4)^2  )
\nonumber\\[0.1cm]
\!\!&=&\!\!\! \lim_{m_q\to 0} w_L^3  (q_3^2, q_4^2 , (q_3+q_4)^2  )
\\
&&\!\!\!\!\!\!
+ \, \frac{F_0 [\stackrel{\rm o}{{\cal F}}_{\pi\gamma^*\gamma^*} (q_3^2,q_4^2) - \stackrel{\rm o}{{\cal F}}_{\pi\gamma^*\gamma^*} (0,0)]}{(q_3 + q_4)^2}  
+ \Delta w^3  (q_3^2, q_4^2 , (q_3+q_4)^2  ) ,
\nonumber
\eea
where $\Delta w^3  (q_3^2, q_4^2 , (q_3+q_4)^2  )$ represents the part of $w_0+w_1$ that is regular 
at $(q_3+q_4)^2=0$ in the chiral limit. The first equality gives the version of Eq. \rf{w0_pi0_pole}
corresponding to the chiral limit. In the
second equality I have isolated the contribution to the pole coming from $\stackrel{\rm o}{{\cal F}}_{\pi\gamma^*\gamma^*} (0,0)$ 
alone, and have identified it, in the third equality, with Eq. \rf{w_L_chiral}. 
Taking now the limit where $q_4$ vanishes or, equivalently,
the combined limit $q_4^2\to 0$ and $(q_3+q_4)^2\to q_3^2$, we see that the relation 
\be
\lim_{q_4\to 0} \lim_{m_q\to 0} \left[ w_0^3 (q_3^2, q_4^2 , (q_3+q_4)^2  ) + w_1^3 (q_3^2, q_4^2 , (q_3+q_4)^2  ) \right]
= \lim_{q_4\to 0} \lim_{m_q\to 0} w_L^3  (q_3^2, q_4^2 , (q_3+q_4)^2  ) = \frac{3{\cal A}/4}{q_3^2}   ,
\ee
which follows from Eq. \rf{wL_q2_vanishes}, rests on an exact cancellation between a contribution that comes from a part of the pion pole,
namely the one involving the momentum dependence of the pion transition form factor, and the contribution that is regular at $(q_3+q_4)^2=0$,
\be
\frac{F_0 [\stackrel{\rm o}{{\cal F}}_{\pi\gamma^*\gamma^*} (q_3^2,0) - \stackrel{\rm o}{{\cal F}}_{\pi\gamma^*\gamma^*} (0,0)]}{q_3^2}  
+ \Delta w^3  (q_3^2, 0 , q_3^2  ) = 0 .
\lbl{cancel}
\ee
The computation in Appendix B shows that this cancellation indeed happens at 
one loop in the low-energy expansion. But it is in fact an exact property of QCD in the chiral limit,
and a direct consequence of the relation \rf{cond_w_i}. Besides its confirmation in the low-energy
expansion, it can also be illustrated in a simple resonance model like the one of Ref. \cite{Kampf:2011ty}.
A straightforward calculation yields
\bea
w_0^3 (q_3^2, q_4^2 , (q_3+q_4)^2  ) + w_1^3 (q_3^2, q_4^2 , (q_3+q_4)^2  ) \!\!&=&\!\!
\frac{1}{(q_3+q_4)^2} \! \left[ \frac{3}{4}\,{\cal A}
+
\frac{b(q_3^2 + q_4^2)}{(q_3^2-M_V^2)(q_4^2-M_V^2)}
+
\frac{c_1 q_3^2 + c_2 q_4^2}{q_3^2-M_V^2} + \frac{c_1 q_4^2 + c_2 q_3^2}{q_4^2-M_V^2} \right]
\nonumber\\
&&\!\!
- \, \frac{b}{(q_3^2-M_V^2)(q_4^2-M_V^2)}
- \frac{c_1}{q_3^2-M_V^2} - \frac{c_2}{q_4^2-M_V^2}   .
\eea
The manner in which the parameters $b$, $c_1$, $c_2$ are related to the resonance couplings
and to the mass $M_V$ of the vector resonance in this model need not concern us here.
What matters instead is to observe that the cancellation \rf{cancel} indeed takes place
when either one of the limit \rf{lim_usual} or \rf{lim_disp} is considered.

Whatever one decides to call the pseudo-paradox \rf{pseudo-paradox} at the origin of the debate in the recent literature,
it rests on a wrong identification, in the chiral limit, of the pion-pole contribution on the right-hand side of
the short-distance constraint in Eq. \rf{SD_W1}, and which itself arises from the identification 
of the two functions $w_L$ and $w_0+w_1$ in the kinematic limit relevant for $a_\mu^{\mbox{\tiny{HLxL}}}$.
This second identification is correct from the functional point of view, but the quite different physical 
contents of these two functions have not been given sufficiently close attention. Once this is done,
the debate loses its \textit{raison d'\^etre}.

\indent

\begin{center}
\textit{3.~~Pseudoscalar poles}
\end{center}

We may now extend the discussion to pseudoscalar poles in general. Strictly speaking,
poles appear only for the lightest of these states, the pseudo-Goldstone mesons $\pi^0$, $\eta$, $\eta'$.
Heavier pseudoscalar states, like for instance the $\pi(1300)$ isotriplet $J^P=0^-$
resonance, are often too broad to be described just as poles on the real axis of the complex
$s$-plane. Such a description 
would require a narrow-width approximation, which finds some justification by
considering, for instance, the large-$N_c$ limit. Let us adopt the latter framework
for the present discussion. In the case of the three-point function ${\cal W}_{\mu\nu\rho}$,
these poles are again to be found in the function $w_0$
\be
w_0 (q_3^2 , q_4^2 , (q_3+q_4)^2) = \sum_P \sum_{a=3,8,0}
{\rm tr}(Q^2 \lambda^a) \frac{F_P^a {\cal F}_{P\gamma^*\gamma^*} (q_3^2,q_4^2)}{(q_3 + q_4)^2 - M_P^2}
+ \cdots  .
\lbl{w0_PS_pole}
\ee
Here the sum runs over all the $J^P=0^-$ states with masses $M_P$, decay constants $F_P^a$,
defined by the matrix elements
\be
\langle \Omega \vert A_\rho^a (0) \vert P(p) \rangle = i F_P^a p_\rho  ,
\ee
and with transitions form factors ${\cal F}_{P\gamma^*\gamma^*}$ defined in 
analogy with the case of the pion in Eq. \rf{pi0_TFF}.

At the level of the four-point function, each of these pseudoscalar states
produces a contribution analogous to the one of the pion,
\be
{\hat{\cal W}}_1^{(P)} (q_1 , q_2 , q_3 , q_4) = 
- \frac{{\cal F}_{P\gamma^*\gamma^*} (q_1^2 , q_2^2 ) \, {\cal F}_{P\gamma^*\gamma^*} (q_3^2 , q_4^2 )}{(q_3 + q_4)^2 - M_P^2}  .
\lbl{W1_P_pole}
\ee
Since the dynamical singularities have to match on both sides of the short-distance
constraint \rf{SD_W1}, we need to check that the relation
\be
{\hat{\cal W}}_1^{(P)} ({\bar q} + {\hat q} , {\bar q} - {\hat q} , q_3 , q_4)
= - \frac{2}{{\hat q}^2} \sum_{a=3,8,0} {\rm tr}(Q^2 \lambda^a) \frac{F_P^a {\cal F}_{P\gamma^*\gamma^*} (q_3^2,q_4^2)}{(q_3 + q_4)^2 - M_P^2}
+ {\cal O} ({\hat q}^{-4})
\ee
holds for asymtotic Euclidian values of the momentum ${\hat q}$. That this is indeed the case 
follows again from Eq. \rf{BJL_limit}. It is thus possible to consider the two limits
discussed previously for the pion-pole contribution to $a_\mu^{\mbox{\tiny{HLxL}}}$.
Without surprise, the outcomes are again different
\be
\lim_{q_4\to 0} \lim_{{\hat q}^2 \to - \infty} {\hat{\cal W}}_1^{(P)} ({\bar q} + {\hat q} , {\bar q} - {\hat q} , q_3 , q_4)
= 
- \frac{2}{{\hat q}^2} \sum_{a=3,8,0} {\rm tr}(Q^2 \lambda^a) \frac{F_P^a {\cal F}_{P\gamma^*\gamma^*} (M_P^2,0)}{q_3^2 - M_P^2} + {\cal O}({\hat q}^{-4}),
\lbl{MV_limit_P}
\ee
\bea
\lim_{\genfrac{}{}{0pt}{}{s\to q_3^2}{q_4^2\to 0}} \lim_{{\hat q}^2 \to - \infty} {\hat{\cal W}}_1^{(P)} ({\bar q} + {\hat q} , {\bar q} - {\hat q} , q_3 , q_4)
=
- \frac{2}{{\hat q}^2} \sum_{a=3,8,0} {\rm tr}(Q^2 \lambda^a) \frac{F_P^a {\cal F}_{P\gamma^*\gamma^*} (q_3^2,0)}{q_3^2 - M_P^2} + {\cal O}({\hat q}^{-4})  .
\lbl{disp-limit_P}
\eea

Finally, in the combined large-$N_c$ and three-flavour chiral limit, 
each one of the flavour-diagonal axial currents defined in Eq. \rf{axial_currents} is conserved,
so that the decay constants vanish as $F_P^a \sim {\cal O}(m_q) + {\cal O}(1/N_c)$
for $P\neq\pi^0 , \eta , \eta'$, and the non-Goldstone pseudoscalar poles in ${\hat{\cal W}}_1$
contribute only to subleading terms of the short-distance expansion.

\section{Summary and conclusion}
\label{sec:summary}
\setcounter{equation}{0}

This note proposes a critical, albeit only partial, discussion of the implications of the 
short-distance constraint of Ref. \cite{Melnikov:2003xd} for 
one of the invariant functions describing the rank-four hadronic vacuum polarization tensor.
This study is focused on a very specific issue under debate in the recent literature,
with the hope that it may contribute positively to this discussion.
To this effect, I have first re-derived the short-distance constraints in the more general 
case of a generic kinematic configuration, and, more importantly, expressed them in terms of functions that are 
free of kinematic singularities. This allows to state a certain number of general properties  
that have to be met and that have been listed after Eq. \rf{SD_W1}.

I have then discussed the two kinematic limits that are currently considered in applications 
to the anomalous magnetic moment of the muon, for both the contribution from the pion pole
or from narrow non-Goldstone pseudoscalar states. 
Working with functions that are free of kinematic 
singularities warrants that both kinematic limits can be taken without problem or ambiguity,
and lead to coherent results if the same limit is taken on both sides of Eq. \rf{SD_W1}.
They are, however, definitely different limits, and as such simply give... different results for the contribution from these poles.
In itself, this needs not necessarily constitute a problem, since the pion pole is but one contribution to $a_\mu^{\mbox{\tiny{HLxL}}}$,
although an important one. But an evaluation of $a_\mu^{\mbox{\tiny{HLxL}}}$ at a level of
precision of $10\%$ in relative terms requires also to include other contributions in a controled manner,
and a comparison between different approaches or prescriptions is only meaningful once this task has been 
completed.

If one follows the evolution of the pion pole through the different limits that are taken,
no ambiguity in its identification arises, and a cancellation mechanism that necessarily
needs to be at work (in QCD) in order to bring the two functions $w_L(q_3^2,0,q_3^2)$ and 
$w_0(q_3^2,0,q_3^2) + w_1(q_3^2,0,q_3^2)$ to an identical form is brought out. 
This mechanism is clearly evidenced in the regime of small momentum transfers, 
where the low-energy expansion can be used. The function that appears on the right-hand side 
of the short-distance constraint is $w_0(q_3^2,0,q_3^2) + w_1(q_3^2,0,q_3^2)$, whose 
pion-pole contribution in the chiral limit is only partially given by the pion pole of
$w_L(q_3^2,0,q_3^2)$.

Narrow pseudoscalar states other than $\pi$, $\eta$, $\eta'$ contribute to both sides of Eq. \rf{SD_W1}
in a perfectly consistent manner. And this consistency persists in the chiral limit, where the 
non-singlet and non-Goldstone pseudoscalar states disappear altogether from the right-hand side
while their contribution to the left-hand side becomes sub-leading in the short-distance expansion. 
If one takes in addition the large-$N_c$ limit, then this situation extends to all non-Goldstone 
pseudoscalar states. 

Finally, let me point out that although I have refered several times to the constraint \rf{SD_W1} or to its 
more general version \rf{MV_SD} as the short-distance condition, the plural form would actually be more appropriate, 
since it really is a constraint on the fourth rank vacuum polarization tensor for each value of the momentum 
transfers $q_3^2$, $q_4^2$ and $(q_3+q_4)^2$. And even in the kinematic regime relevant for the evaluation of 
$a_\mu^{\mbox{\tiny{HLxL}}}$ it still gives a condition for each value of $q_3^2$ and not only when $q_3^2$ becomes 
large in the Euclidian region, as it is most of the time being used. No phenomenological approach or model designed for the 
evaluation of $a_\mu^{\mbox{\tiny{HLxL}}}$ I am aware of has, so far, exploited the full content of 
the condition of Ref. \cite{Melnikov:2003xd} in this broader sense.

\indent

\begin{center}
 {\bf Acknowledgements}
\end{center}
 I wish to thank E. de Rafael for a careful reading of the manuscript, for useful suggestions,
 and for many stimulating and informative discussions about g-2 related issues over the years.

\indent

\appendix

\setcounter{section}{0}

\def\theequation{\Alph{section}.\arabic{equation}}

\begin{center}
{\large\textbf{Appendix A}}
\end{center}
\def\theequation{A.\arabic{equation}}
\setcounter{equation}{0}

In this appendix, I describe how the short-distance constraint given in Eq. \rf{MV_SD_ter}
for the four-point function ${\cal W}^{\mu\nu\rho\sigma}$ can be transformed into short-distance
constraints for some of the individual invariant functions ${\cal W}_i$ introduced in Eq. \rf{W4_decomp}.
This is a somewhat lengthy process, so that only a brief outline of it will be presented.
Before that, I first give the dictionnary between the notation used here and the one used in 
Refs. \cite{Colangelo:2015ama,Colangelo:2017fiz,Colangelo:2019lpu,Colangelo:2019uex}.

The tensor 
$\Pi^{\mu\nu\rho\sigma} (q_1 , q_2 , q_3 , q_4)$ defined in Eq. (3.1) of \cite{Colangelo:2015ama}
is related to ${\cal W}^{\mu\nu\rho\sigma} (q_1 , q_2 , q_3 , q_4)$ by
\be
{\cal W}^{\mu\nu\rho\sigma} (q_1 , q_2 , q_3 , q_4) = \Pi^{\mu\nu\rho\sigma} (-q_1 , -q_2 , -q_3 , q_4) 
 = \Pi^{\mu\nu\rho\sigma} (q_1 , q_2 , q_3 , -q_4)
.
\lbl{conventions}
\ee
The absence of a minus sign in the last entry of $\Pi^{\mu\nu\rho\sigma}$ in the 
first equality is due to the fact that in Refs. 
\cite{Colangelo:2015ama,Colangelo:2017fiz} the momenta $q_1$, $q_2$
and $q_3$ are taken as incoming, whereas $q_4$ is taken as outgoing.
The second equality follows from the fact that the tensors $\Pi^{\mu\nu\rho\sigma} (q_1 , q_2 , q_3 , q_4)$
or ${\cal W}^{\mu\nu\rho\sigma} (q_1 , q_2 , q_3 , q_4)$ remain the same if all
momenta are reversed simultaneously. The tensors $T_i^{\mu\nu\rho\sigma}$ are listed in Eq. (3.14) and in Appendix 
B of Ref. \cite{Colangelo:2015ama}. I have taken the ``seed tensors'' displayed in  Eq. (3.14) of Ref. \cite{Colangelo:2015ama}
as they stand, i.e. without changing the sign of $q_4$, and have then applied the symmetry operations
listed in Eq. (B.1) of Ref. \cite{Colangelo:2015ama}, with the difference that the exchange operation
${\cal C}_{14}$, for instance, means $(\mu , q_1) \leftrightarrow (\sigma , q_4)$, i.e. without changing 
the sign of $q_4$. The relations between the invariant functions then read $\Pi_i (q_1 , q_2 , q_3 , q_4) =
\pm{\cal W}_i (q_1 , q_2 , q_3 , q_4)$ and it is easy to trace which sign applies for a specific value of $i$.
This explains, for instance, why in the definition of ${\hat W}_1$ in Eq. \rf{W1hat} there is a relative
minus sign between the two terms, whereas one finds a plus sign in Eq. (2.15) of Ref. \cite{Colangelo:2017fiz}.

Coming back to Eq. \rf{MV_SD_ter}, each function ${\cal W}_i$ has, in the limit under consideration, an expansion of the form
\be
 {\cal W}_i ({\bar q} + {\hat q},{\bar q} - {\hat q},q_3 , q_4)
= \frac{1}{{\hat q}^{n_i}} \left[ {\cal W}_i^{[n_i]} (q_3 , q_4) + \frac{{\hat q}^\mu}{{\hat q}^2} \, {\cal W}_{i,\mu}^{[n_i + 1]} (q_3 , q_4)
+ \cdots \right]
.
\lbl{Wi_expand}
\ee
The value of $n_i$, which determines the leading power behaviour, can be fixed
in the following manner: the tensors $T_i^{\mu\nu\rho\sigma}$
have dimension 4 for $i=1,\ldots , 6$, dimension 8 for $i=31, \ldots , 36$, and
dimension 6 in all other cases, whereas the tensor ${\cal W}^{\mu\nu\rho\sigma}$
is dimensionless. Furthermore, we are looking for relations of the type
\be
{\cal W}_i^{[n_i]} (q_3 , q_4) = \sum_{k=0}^3 c_{ik} w_k (q_3, q_4)  ,
\ee
with some numerical coefficients $c_{ik}$, and where the functions $w_k (q_3 , q_4)$ 
have dimension $-2$. This means that one has $n_i = 2$ for $i=1,\ldots , 6$,
$n_i = 6$ for $i=31,\ldots , 36$, and $n_i = 4$ for the remaining values of $i$.
It is then possible to proceed upon going through the following steps:
\begin{itemize}
\item 
First, one notices that the highest power in ${\hat q}$ of each tensor 
$T_i^{\mu\nu\rho\sigma} ({\bar q} + {\hat q},{\bar q} - {\hat q}, q_3 , q_4)$, which
is given by 
\be
{\widehat T}_i^{\mu\nu\rho\sigma} ({\hat q}, q_3, q_4) \equiv T_i^{\mu\nu\rho\sigma} ({\hat q}, - {\hat q}, q_3, q_4) ,
\ee
also varies from case to case. This highest power is simply equal to $1$ for $i=1$,
to $2$ for $i=2,\ldots , 6,9,12,13,15,17,$
$18,29,\-30,32,37,48,49$, and so on.
Since we are looking for a behaviour
that does not decrease faster than $1/{\hat q}$ when ${\hat q}$ becomes large,
we are eventually left with only the cases
\be
i = 1,\ldots , 8,10,11,14,16,19, \ldots ,28,31,38, \ldots 47,50, \ldots 54
\ee
to consider.
\item
Second, for each of these cases, one extracts from the tensor $T_i^{\mu\nu\rho\sigma} ({\bar q} + {\hat q},{\bar q} - {\hat q}, q_3 , q_4)$
the part, denoted as  ${\cal T}_i^{\mu\nu\rho\sigma} ({\hat q}, q_3 , q_4)$, that is either linear in ${\hat q}$, i.e.
\be
{\cal T}_i^{\mu\nu\rho\sigma} ({\hat q}, q_3 , q_4) = T_i^{\mu\nu\rho\sigma} ({\hat q}, {\bar q}, q_3, q_4)
+ T_i^{\mu\nu\rho\sigma} ({\bar q}, - {\hat q}, q_3, q_4)  ,
\ee
for $i=2, \ldots ,6$, or that is of the form ${\hat q}^2$ times terms linear in ${\hat q}$, when it exists,
in the other cases, except $i=1$, where one has
\be
{\cal T}_1^{\mu\nu\rho\sigma} ({\hat q}, q_3 , q_4) =
T_1^{\mu\nu\rho\sigma} ({\bar q} + {\hat q},{\bar q} - {\hat q}, q_3 , q_4) =
K_0^{\mu\nu\rho\sigma} ({\hat q} , q_3 , q_4)
.
\ee 
With these pieces at hand, one can then construct a set of other useful relations involving the 
tensors $K_i^{\mu\nu\rho\sigma} ({\hat q} , q_3 , q_4)$ defined in Eq. \rf{K-tensors}:
\be
K_0^{\mu\nu\rho\sigma} ({\hat q} , q_3 , q_4) =
\frac{1}{{\hat q}^2} \left[ {\cal T}_{46}^{\mu\nu\rho\sigma} ({\hat q}, q_3 , q_4) + {\cal T}_{47}^{\mu\nu\rho\sigma} ({\hat q}, q_3 , q_4)\right]
- 2 \left[ {\cal T}_5^{\mu\nu\rho\sigma} ({\hat q}, q_3 , q_4) + {\cal T}_6^{\mu\nu\rho\sigma} ({\hat q}, q_3 , q_4) \right]   ,
\ee
\bea
\nonumber\\
K_3^{\mu\nu\rho\sigma} ({\hat q} , q_3 , q_4) - \frac{1}{2} K_1^{\mu\nu\rho\sigma} ({\hat q} , q_3 , q_4) \!\!&=&\!\!
\frac{1}{{\hat q}^2}  \left[ x_1 {\cal T}_{10}^{\mu\nu\rho\sigma} ({\hat q}, q_3 , q_4) - y_1 {\cal T}_{11}^{\mu\nu\rho\sigma} ({\hat q}, q_3 , q_4)
\right. 
\nonumber\\ 
&&
\left.
\ \ + \, (1-y_1) {\cal T}_{14}^{\mu\nu\rho\sigma} ({\hat q}, q_3 , q_4) - (1-x_1) {\cal T}_{16}^{\mu\nu\rho\sigma} ({\hat q}, q_3 , q_4)
\right] 
\nonumber\\
&&\hspace{-0.55cm}
+ \, \frac{1}{2{\hat q}^2} 
\left[ {\cal T}_{46}^{\mu\nu\rho\sigma} ({\hat q}, q_3 , q_4) - {\cal T}_{47}^{\mu\nu\rho\sigma} ({\hat q}, q_3 , q_4) \right]   ,
\eea
\bea
\nonumber\\
K_2^{\mu\nu\rho\sigma} ({\hat q} , q_3 , q_4)  \!\!&=&\!\!
\frac{x_2}{{\hat q}^2}  \left[ {\cal T}_{50}^{\mu\nu\rho\sigma} ({\hat q}, q_3 , q_4) + {\cal T}_{51}^{\mu\nu\rho\sigma} ({\hat q}, q_3 , q_4)
+ {\cal T}_{52}^{\mu\nu\rho\sigma} ({\hat q}, q_3 , q_4) + {\cal T}_{53}^{\mu\nu\rho\sigma} ({\hat q}, q_3 , q_4)
\right] 
\nonumber\\
&&\hspace{-0.55cm}
+ \, \frac{1-x_2}{{\hat q}^2} 
\left[ y_2 \left( {\cal T}_{38}^{\mu\nu\rho\sigma} ({\hat q}, q_3 , q_4) + {\cal T}_{39}^{\mu\nu\rho\sigma} ({\hat q}, q_3 , q_4) \right)
\right. 
\nonumber\\ 
&&
\left.
\hspace{0.45cm} 
+ \, (1-y_2) \left( {\cal T}_{40}^{\mu\nu\rho\sigma} ({\hat q}, q_3 , q_4) + {\cal T}_{41}^{\mu\nu\rho\sigma} ({\hat q}, q_3 , q_4) \right)
\right. 
\\ 
&&
\left.
\hspace{0.45cm} 
- \, {\cal T}_{21}^{\mu\nu\rho\sigma} ({\hat q}, q_3 , q_4) - {\cal T}_{23}^{\mu\nu\rho\sigma} ({\hat q}, q_3 , q_4)
- {\cal T}_{25}^{\mu\nu\rho\sigma} ({\hat q}, q_3 , q_4) - {\cal T}_{27}^{\mu\nu\rho\sigma} ({\hat q}, q_3 , q_4)
\right]   ,
\nonumber
\eea
\bea
\nonumber\\
K_3^{\mu\nu\rho\sigma} ({\hat q} , q_3 , q_4)  \!\!&=&\!\!
\frac{x_3}{{\hat q}^2}  \left[ {\cal T}_{50}^{\mu\nu\rho\sigma} ({\hat q}, q_3 , q_4) + {\cal T}_{51}^{\mu\nu\rho\sigma} ({\hat q}, q_3 , q_4)
- {\cal T}_{52}^{\mu\nu\rho\sigma} ({\hat q}, q_3 , q_4) - {\cal T}_{53}^{\mu\nu\rho\sigma} ({\hat q}, q_3 , q_4)
\right] 
\nonumber\\
&&\hspace{-0.55cm}
- \, \frac{1-x_3}{{\hat q}^2} 
\left[ y_3 \left( {\cal T}_{38}^{\mu\nu\rho\sigma} ({\hat q}, q_3 , q_4) - {\cal T}_{39}^{\mu\nu\rho\sigma} ({\hat q}, q_3 , q_4) \right)
\right. 
\nonumber\\ 
&&
\left.
\hspace{0.45cm} 
- \, (1-y_3) \left( {\cal T}_{40}^{\mu\nu\rho\sigma} ({\hat q}, q_3 , q_4) - {\cal T}_{41}^{\mu\nu\rho\sigma} ({\hat q}, q_3 , q_4) \right)
\right. 
\\ 
&&
\left.
\hspace{0.45cm} 
+ \, {\cal T}_{21}^{\mu\nu\rho\sigma} ({\hat q}, q_3 , q_4) - {\cal T}_{23}^{\mu\nu\rho\sigma} ({\hat q}, q_3 , q_4)
+ {\cal T}_{25}^{\mu\nu\rho\sigma} ({\hat q}, q_3 , q_4) - {\cal T}_{27}^{\mu\nu\rho\sigma} ({\hat q}, q_3 , q_4)
\right]   .
\nonumber
\eea
In these identities $x_{1,2,3}$ and $y_{1,2,3}$ are real parameters belonging to the interval $[0,1]$,
but can otherwise be chosen arbitrarily. 
To these, one also has to add the two following relations:
\bea 
\frac{{\hat q}^2}{2} \left[ K_2^{\mu\nu\rho\sigma} ({\hat q} , q_3 , q_4) + K_3^{\mu\nu\rho\sigma} ({\hat q} , q_3 , q_4) \right]  
\!\!&=&\!\!
{\widehat T}_{42}^{\mu\nu\rho\sigma} ({\hat q}, q_3, q_4) + {\widehat T}_{43}^{\mu\nu\rho\sigma} ({\hat q}, q_3, q_4)
\nonumber\\
&&\!\!
+ \, ({\hat q} \cdot q_3) \left[
{\widehat T}_{2}^{\mu\nu\rho\sigma} ({\hat q}, q_3, q_4) - {\widehat T}_{3}^{\mu\nu\rho\sigma} ({\hat q}, q_3, q_4)
\right.
\nonumber\\
&&\!\!
\left.
\qquad\qquad
- \, {\widehat T}_{5}^{\mu\nu\rho\sigma} ({\hat q}, q_3, q_4) + {\widehat T}_{6}^{\mu\nu\rho\sigma} ({\hat q}, q_3, q_4)
\right] \!,~~~~~~~
\eea
and
\bea 
\frac{{\hat q}^2}{2} \left[ K_2^{\mu\nu\rho\sigma} ({\hat q} , q_3 , q_4) - K_3^{\mu\nu\rho\sigma} ({\hat q} , q_3 , q_4) \right]  
\!\!&=&\!\!
{\widehat T}_{44}^{\mu\nu\rho\sigma} ({\hat q}, q_3, q_4) + {\widehat T}_{45}^{\mu\nu\rho\sigma} ({\hat q}, q_3, q_4)
\nonumber\\
&&\!\!
- \, ({\hat q} \cdot q_4) \left[
{\widehat T}_{2}^{\mu\nu\rho\sigma} ({\hat q}, q_3, q_4) - {\widehat T}_{3}^{\mu\nu\rho\sigma} ({\hat q}, q_3, q_4)
\right.
\nonumber\\
&&\!\!
\left.
\qquad\qquad
- \, {\widehat T}_{5}^{\mu\nu\rho\sigma} ({\hat q}, q_3, q_4) + {\widehat T}_{6}^{\mu\nu\rho\sigma} ({\hat q}, q_3, q_4)
\right] \!.~~~~~~~
\eea
\item
Next, one expands the fonctions ${\cal W}_i$ as explained in Eq. \rf{Wi_expand}, 
taking into account the symmetry properties of these functions that are listed in 
Ref. \cite{Colangelo:2015ama}. It then remains to collect in the four-point function
all the terms that do not decrease faster than $1/{\hat q}$ and to require that their
sum matches the right-hand side of Eq. \rf{MV_SD_ter}. The result of this exercise 
then leads to the following relations:
\bea 
{\cal W}_1^{[2]} (q_3 , q_4) + \frac{1}{2} \left[ {\cal W}_{46}^{[4]} (q_3 , q_4) + {\cal W}_{46}^{[4]} (q_4 , q_3) \right] \!\!&=&\!\!
- \, 2 w_0 (q_3 , q_4)  ,
\nonumber\\[0.25cm]
\frac{1}{2} \left[ {\cal W}_{46}^{[4]} (q_3 , q_4) - {\cal W}_{46}^{[4]} (q_4 , q_3) \right] \!\!&=&\!\!
+ \, 2 w_1 (q_3 , q_4)  ,
\nonumber\\[0.25cm]
\left[ {\cal W}_{38}^{[4]} (q_3 , q_4) + {\cal W}_{38}^{[4]} (q_4 , q_3) \right] \!\!&&
\nonumber\\
+ \, \frac{1}{2} \left[ {\cal W}_{42}^{[4]} (q_3 , q_4) + {\cal W}_{42}^{[4]} (q_4 , q_3) 
+ {\cal W}_{50}^{[4]} (q_3 , q_4) + {\cal W}_{50}^{[4]} (q_4 , q_3) \right] \!\!&=&\!\!
- \, 2 w_2 (q_3 , q_4)  ,
\nonumber\\[0.25cm]
\left[ {\cal W}_{46}^{[4]} (q_3 , q_4) - {\cal W}_{46}^{[4]} (q_4 , q_3) \right] -
\left[ {\cal W}_{38}^{[4]} (q_3 , q_4) - {\cal W}_{38}^{[4]} (q_4 , q_3) \right] \!\!&&
\nonumber\\
+ \, \frac{1}{2} \left[ {\cal W}_{42}^{[4]} (q_3 , q_4) - {\cal W}_{42}^{[4]} (q_4 , q_3) 
+ {\cal W}_{50}^{[4]} (q_3 , q_4) - {\cal W}_{50}^{[4]} (q_4 , q_3) \right] \!\!&=&\!\!
- \, 2 w_3 (q_3 , q_4)  ,
\eea
together with
\be
{\cal W}_2^{[2]} (q_3 , q_4) = {\cal W}_3^{[2]} (q_3 , q_4) = {\cal W}_4^{[2]} (q_3 , q_4) = 0  ,
\quad 
{\cal W}_{2,\mu}^{[3]} (q_3 , q_4) - q_{3\mu} {\cal W}_{42}^{[4]} (q_3 , q_4) + q_{4\mu} {\cal W}_{42}^{[4]} (q_4 , q_3) = 0  ,
\ee
\be 
{\cal W}_5^{[2]} (q_3 , q_4) + {\cal W}_{10}^{[4]} (q_3 , q_4) + {\cal W}_{10}^{[4]} (q_4 , q_3) = 0  ,
\quad
{\cal W}_{10}^{[4]} (q_3 , q_4) = {\cal W}_{46}^{[4]} (q_3 , q_4)  ,
\ee
\bea
{\cal W}_{5,\mu}^{[3]} (q_3 , q_4) + {\cal W}_{10,\mu}^{[5]} (q_3 , q_4) - {\cal W}_{10,\mu}^{[5]} (q_4 , q_3)
+ q_{3\mu} \left[ {\cal W}_{22}^{[4]} (q_3 , q_4) + {\cal W}_{42}^{[4]} (q_3 , q_4)  \right] \!\!&&
\nonumber\\
- q_{4\mu} \left[ {\cal W}_{22}^{[4]} (q_4 , q_3) + {\cal W}_{42}^{[4]} (q_4 , q_3)  \right]
- {\cal W}_{54,\mu}^{[5]} (q_3 , q_4)
\!\!&=&\!\! 0  , ~~~
\eea
\be
{\cal W}_{7}^{[4]} (q_3 , q_4) = {\cal W}_{19}^{[4]} (q_3 , q_4)  ,
\quad 
{\cal W}_{21}^{[4]} (q_3 , q_4) = - 2 {\cal W}_{38}^{[4]} (q_4 , q_3)  .
\ee
\end{itemize}
Introducing, along Eqs. (2.15) and (2.16) of Ref. \cite{Colangelo:2017fiz}, the 
functions ${\hat{\cal W}}_i$, with the appropriate changes of signs due to 
the differences in the conventions, as discussed after Eq. \rf{conventions} above,
one then establishes Eq. \rf{SD_W1} and, for instance, 
\bea
{\hat{\cal W}}_5 ({\bar q} + {\hat q},{\bar q} - {\hat q}, q_3 , q_4)
+ {\hat{\cal W}}_6 ({\bar q} + {\hat q},{\bar q} - {\hat q}, q_3 , q_4)
- {\hat q}^2 \left[ 
{\hat{\cal W}}_{50} ({\bar q} + {\hat q},{\bar q} - {\hat q}, q_3 , q_4) + {\hat{\cal W}}_{51} ({\bar q} + {\hat q},{\bar q} - {\hat q}, q_3 , q_4) 
\right]
= &&
\nonumber\\
&& \hspace{-10.5cm}
= \, \frac{4}{{\hat q}^2} \left[ w_2 (q_3^2,q_4^2,(q_3+q_4)^2) + w_3 (q_3^2,q_4^2,(q_3+q_4)^2) \right] + {\cal O} ({\hat q}^{-4}) ,
\eea
or
\bea
{\hat{\cal W}}_{50} ({\bar q} + {\hat q},{\bar q} - {\hat q}, q_3 , q_4) + {\hat{\cal W}}_{51} ({\bar q} + {\hat q},{\bar q} - {\hat q}, q_3 , q_4) 
+ 2 {\hat{\cal W}}_{39} ({\bar q} + {\hat q},{\bar q} - {\hat q}, q_3 , q_4) = &&
\nonumber\\
&& \hspace{-7.0cm}
=  - \frac{4}{{\hat q}^4} \! \left[ w_2 (q_3^2,q_4^2,(q_3+q_4)^2) + w_3 (q_3^2,q_4^2,(q_3+q_4)^2) \right] + {\cal O} ({\hat q}^{-6}) .
\eea
The limit $q_4 \to 0$ of these two last relations can also be recovered from 
the expressions given in Eq. (3.25) of Ref. \cite{Colangelo:2019uex}.

\indent

\begin{center}
{\large\textbf{Appendix B}}
\end{center}
\def\theequation{B.\arabic{equation}}
\setcounter{equation}{0}


In this appendix I reproduce the expressions of the invariant functions $w_i$, $i=0,1,2,3$,
which provide a decomposition of the three-point function ${\cal W}^{\mu\nu\rho}$, 
obtained from a one-loop calculation in the low-energy expansion \cite{Weinberg:1978kz,Gasser:1983ky,Gasser:1983yg}
with three light flavours \cite{Gasser:1984gg}. For this, I also need the 
effective lagrangian at order ${\cal O}(p^6)$ in the sector of odd intrinsic parity,
whose general structure has been worked out in Refs. \cite{Ebertshauser:2001nj,Bijnens:2001bb}.
For definiteness, I will adopt the basis of counterterms given in the second of these two references.
These expressions then allow to discuss, within this framework, a certain number of properties 
mentioned at various places in the main text. For reasons of simplicity, I only give the 
expressions corresponding to the iso-triplet channel. Similar expressions can be worked out 
in the octet channel. A discussion of the singlet channel would require to work within the framework
of a combined chiral and $1/N_c$ expansion, which is in principle also possible, since the 
necessary tools are available \cite{Gasser:1984gg,HerreraSiklody:1996pm,Kaiser:2000gs}.

At one loop in chiral perturbation theory, one obtains the following results
\bea
w_1^3 (q_1^2 , q_2^2 , (q_1+q_2)^2) \!\!&=&\!\! \frac{N_c}{144 \pi^2 F_\pi^2}
\left[\left(1 - \frac{4 M_\pi^2}{q_1^2} \right) {\bar J}_{\pi\pi} (q_1^2)  - \left(1 - \frac{4 M_\pi^2}{q_2^2} \right) {\bar J}_{\pi\pi} (q_2^2) \right]
\nonumber\\
&&\hspace{-0.5cm}
+ \, \frac{N_c}{144 \pi^2 F_\pi^2}
\left[\left(1 - \frac{4 M_K^2}{q_1^2} \right) {\bar J}_{KK} (q_1^2)  - \left(1 - \frac{4 M_K^2}{q_2^2} \right) {\bar J}_{KK} (q_2^2) \right]
+ {\cal O}(p^8)    ,
\nonumber\\ 
w_2^3 (q_1^2 , q_2^2 , (q_1+q_2)^2) \!\!&=&\!\! - \, \frac{16}{3} C_{22}^W (\mu) 
+ \, \frac{N_c}{72 \pi^2 F_\pi^2} \frac{1}{16 \pi^2} \left( \ln\frac{M_\pi^2}{\mu^2} + \ln\frac{M_K^2}{\mu^2}  + \frac{2}{3} \right)
\nonumber\\
&&\hspace{-0.5cm}
- \, \frac{N_c}{144 \pi^2 F_\pi^2}
\left[\left(1 - \frac{4 M_\pi^2}{q_1^2} \right) {\bar J}_{\pi\pi} (q_1^2)  + \left(1 - \frac{4 M_\pi^2}{q_2^2} \right) {\bar J}_{\pi\pi} (q_2^2) \right]   
\nonumber\\
&&\hspace{-0.5cm}
- \, \frac{N_c}{144 \pi^2 F_\pi^2}
\left[\left(1 - \frac{4 M_K^2}{q_1^2} \right) {\bar J}_{KK} (q_1^2)  + \left(1 - \frac{4 M_K^2}{q_2^2} \right) {\bar J}_{KK} (q_2^2) \right]
+ {\cal O}(p^8)    ,
\nonumber\\ 
w_3^3 (q_1^2 , q_2^2 , (q_1+q_2)^2) \!\!&=&\!\! - w_1^3 (q_1^2 , q_2^2 , (q_1+q_2)^2)   
+ {\cal O}(p^8)    ,
\eea
and
\bea
{\cal H}^3 (q_1^2 , q_2^2 , (q_1+q_2)^2) \!\!&=&\!\! \frac{64}{3} C_7^W M_\pi^2 + \frac{M_\pi^2 F_\pi {\widehat{\cal F}}_{\pi\gamma^*\gamma^*} (q_1^2,q_2^2)}{(q_1 + q_2)^2 - M_\pi^2}
+ {\cal O}(p^8)    ,
\nonumber\\
F_\pi {\widehat{\cal F}}_{\pi\gamma^*\gamma^*} (q_1^2,q_2^2) \!\!&=&\!\! {\cal A}^3 + \frac{64}{3} C_7^W M_\pi^2  
\nonumber\\
&&\hspace{-0.5cm}
+ \left[ - \frac{16}{3}   C_{22}^{W} (\mu) 
+ \, \frac{N_c}{72 \pi^2 F_\pi^2} \frac{1}{16 \pi^2} \left( \ln\frac{M_\pi^2}{\mu^2} + \ln\frac{M_K^2}{\mu^2}  + \frac{2}{3} \right) \right] (q_1^2 + q_2^2)
\nonumber\\
&&\hspace{-0.5cm}
- \, \frac{N_c}{72 \pi^2 F_\pi^2}
\left[ (q_1^2 - 4 M_\pi^2) {\bar J}_{\pi\pi} (q_1^2)  + (q_2^2 - 4 M_\pi^2) {\bar J}_{\pi\pi} (q_2^2) \right]   
\nonumber\\
&&\hspace{-0.5cm}
- \, \frac{N_c}{72 \pi^2 F_\pi^2}
\left[ (q_1^2 - 4 M_\pi^2) {\bar J}_{KK} (q_1^2)  + (q_2^2 - 4 M_\pi^2) {\bar J}_{KK} (q_2^2) \right]   
+ {\cal O}(p^8)    ,
\eea
where ${\cal A}^3 = -N_c/12\pi^2 = (3/4){\cal A}$ and the loop function ${\bar J}_{PP}$,
$P=\pi,K$, is defined in Ref. \cite{Gasser:1984gg} and can be conveniently expressed as
the integral
\be
{\bar J}_{PP} (s) = - \frac{1}{16 \pi^2} \int_0^1 du \ln \Big[1 - \frac{s}{M_P^2} u (1-u)\Big] .
\ee
Furthermore, $\mu$ denotes the chiral renormalization scale. The low-energy constant $C_7^W$ is $\mu$-independent,
while the $\mu$-dependence of the renormalized constant $C_{22}^W (\mu)$ is compensated
by the $\log \mu^2$ terms, see Ref.  \cite{Bijnens:2001bb}.
Notice that despite the suggestive notation, and as the $\,{\widehat{ }}\,$ symbol is meant to
remind of, ${\widehat{\cal F}}_{\pi\gamma^*\gamma^*} (q_1^2,q_2^2)$ is not yet the  
pion transition form factor ${\cal F}_{\pi\gamma^*\gamma^*} (q_1^2,q_2^2)$. The relation between 
the two is given by
\be
{\cal F}_{\pi\gamma^*\gamma^*} ((p/2+q)^2,(p/2-q)^2) = \lim_{p^2 \to M_\pi^2} 
{\widehat{\cal F}}_{\pi\gamma^*\gamma^*} ((p/2+q)^2,(p/2-q)^2) .
\ee
In the semi-off-shell case the expression of ${\cal F}_{\pi\gamma^*\gamma^*} ((p/2\pm q)^2,0)$
one obtains this way reproduces the one that is given in Ref. \cite{Bijnens:1988kx}.
From these formulas, one deduces, through the relations given in Eq. \rf{cond_w_i}, the one-loop expression of the remaining functions
\be
w_0^3 (q_1^2 , q_2^2 , (q_1+q_2)^2) = - w_2^3 (q_1^2 , q_2^2 , (q_1+q_2)^2) + \frac{F_\pi {\widehat{\cal F}}_{\pi\gamma^*\gamma^*} (q_1^2,q_2^2)}{(q_1 + q_2)^2 - M_\pi^2}   
+ {\cal O}(p^8)    ,
\ee
and
\bea
\frac{1}{8\pi^2} \, w_L^3 (q_1^2 , q_2^2 , (q_1+q_2)^2) &=& \frac{1}{(q_1+q_2)^2}
\left[ {\cal A}^3 + \frac{64}{3} C_7^W M_\pi^2 + \frac{M_\pi^2 F_\pi {\widehat{\cal F}}_{\pi\gamma^*\gamma^*} (q_1^2,q_2^2)}{(q_1 + q_2)^2 - M_\pi^2} \right]
+ {\cal O}(p^8)    
\nonumber\\
&=& \frac{1}{(q_1+q_2)^2}
\left[ F_\pi {\widehat{\cal F}}_{\pi\gamma^*\gamma^*} (0,0) + \frac{M_\pi^2 F_\pi {\widehat{\cal F}}_{\pi\gamma^*\gamma^*} (q_1^2,q_2^2)}{(q_1 + q_2)^2 - M_\pi^2} \right]
+ {\cal O}(p^8)    
\nonumber\\
&=& \frac{F_\pi {\widehat{\cal F}}_{\pi\gamma^*\gamma^*} (0,0)}{(q_1 + q_2)^2 - M_\pi^2} 
+ \frac{M_\pi^2}{(q_1+q_2)^2}
\, \frac{ F_\pi \left[ {\widehat{\cal F}}_{\pi\gamma^*\gamma^*} (q_1^2,q_2^2) - {\widehat{\cal F}}_{\pi\gamma^*\gamma^*} (0,0) \right]}{(q_1 + q_2)^2 - M_\pi^2} 
+ {\cal O}(p^8)  .~~~~~~
\eea
The kinematic singularity, at $(q_1+q_2)^2=0$, of $w_L^3$ is immediately visible in
this expression. It also shows how, in the chiral limit, this kinematic singularity 
transforms into a dynamical singularity due to the massless pion pole, but with a constant residue, 
fixed by the anomaly,
\be
\lim_{m_q \to 0} F_\pi {\widehat{\cal F}}_{\pi\gamma^*\gamma^*} (0,0) = {\cal A}^3
\ee
The combination that appears in the short-distance condition \rf{SD_W1}
for ${\hat{\cal W}}_1$ is completely different already at one loop, since 
\bea
w_0^3 (q_1^2 , q_2^2 , (q_1+q_2)^2) + w_1^3 (q_1^2 , q_2^2 , (q_1+q_2)^2)
\!\!&=&\!\! 
\frac{F_\pi {\widehat{\cal F}}_{\pi\gamma^*\gamma^*} (q_1^2,q_2^2)}{(q_1 + q_2)^2 - M_\pi^2} 
\nonumber\\
&&\!\!\!
+ \, \frac{16}{3} C_{22}^{W} (\mu) 
- \frac{N_c}{72 \pi^2 F_\pi^2} \frac{1}{16 \pi^2}  \left( \ln\frac{M_\pi^2}{\mu^2} + \ln\frac{M_K^2}{\mu^2}  + \frac{2}{3}  \right)
\nonumber\\
&&\!\!\!
+ \, \frac{N_c}{72 \pi^2 F_\pi^2} \left(1 - \frac{4 M_\pi^2}{q_1^2} \right) {\bar J}_{\pi\pi} (q_1^2)
+  \frac{N_c}{72 \pi^2 F_\pi^2} \left(1 - \frac{4 M_K^2}{q_1^2} \right) {\bar J}_{KK} (q_1^2) ~~~~~
\nonumber\\
&&\!\!\!
+ {\cal O}(p^8)   .
\eea 
As stated in the text, it exhibits a pion pole, with residue given by $F_\pi {\cal F}_{\pi\gamma^*\gamma^*} (q_1^2,q_2^2)$
that retains a non-trivial momentum dependence even in the chiral limit.
The difference between the two expressions can be given a suggestive form,
\bea
&&
w_0^3 (q_1^2 , q_2^2 , (q_1+q_2)^2) + w_1^3 (q_1^2 , q_2^2 , (q_1+q_2)^2) - \frac{1}{8\pi^2} \, w_L^3 (q_1^2 , q_2^2 , (q_1+q_2)^2) 
\nonumber\\
&&
= \left[ 1 - \frac{q_1^2}{(q_1+q_2)^2} \right] \bigg\{ \frac{16}{3} C_{22}^{W} (\mu) 
- \frac{N_c}{72 \pi^2 F_\pi^2} \frac{1}{16 \pi^2}  \left( \ln\frac{M_\pi^2}{\mu^2} + \ln\frac{M_K^2}{\mu^2}  + \frac{2}{3}  \right)
\nonumber\\
&&
\qquad + \, \frac{N_c}{72 \pi^2 F_\pi^2} \left(1 - \frac{4 M_\pi^2}{q_1^2} \right) {\bar J}_{\pi\pi} (q_1^2)
+  \frac{N_c}{72 \pi^2 F_\pi^2} \left(1 - \frac{4 M_K^2}{q_1^2} \right) {\bar J}_{KK} (q_1^2) \bigg\}
\nonumber\\
&&
- \, \frac{q_2^2}{(q_1+q_2)^2}\bigg\{ \frac{16}{3} C_{22}^{W} (\mu) 
- \frac{N_c}{72 \pi^2 F_\pi^2} \frac{1}{16 \pi^2}  \left( \ln\frac{M_\pi^2}{\mu^2} + \ln\frac{M_K^2}{\mu^2}  + \frac{2}{3}  \right)
\nonumber\\
&&
\qquad + \, \frac{N_c}{72 \pi^2 F_\pi^2} \left(1 - \frac{4 M_\pi^2}{q_2^2} \right) {\bar J}_{\pi\pi} (q_2^2)
+  \frac{N_c}{72 \pi^2 F_\pi^2} \left(1 - \frac{4 M_K^2}{q_2^2} \right) {\bar J}_{KK} (q_2^2) \bigg\}
+ {\cal O}(p^8)   .
\eea
It clearly exhibits the cancellation that takes place in the limit $q_2\to 0$.
The corresponding expressions in the chiral limit $m_q\to 0$ can be easily worked out from the 
formulas given above, using
\be
\lim_{M_P \to 0} \left[ {\bar J}_{PP} (s) - \frac{1}{16\pi^2} \ln \frac{M_P^2}{\mu^2}  \right] = - \frac{1}{16\pi^2} \ln \frac{-s}{\mu^2} + \frac{1}{8\pi^2} .
\ee 
Whether one then takes the limit where the four-vector $q_2$ vanishes, 
or the combined, ``dispersive-friendly'', limit $q_2^2\to 0$, $(q_1+q_2)^2\to q_1^2$,
one obtains the same result,
\be
\lim_{q_2\to 0}
\lim_{m_q \to 0} \left[ 
w_0^3 (q_1^2 , q_2^2 , (q_1+q_2)^2) + w_1^3 (q_1^2 , q_2^2 , (q_1+q_2)^2) 
\right]
= \!\!
\lim_{q_2\to 0}
\lim_{m_q \to 0} \frac{1}{8 \pi^2} \, w_L^3 (q_1^2 , q_2^2 , (q_1+q_2)^2)
= \frac{{\cal A}^3}{q_1^2}
.
\lbl{limit_q2_to_0}
\ee
But the manner how this result comes about is totally different in the two cases. 
To see this in an easy manner, let me consider the 
combined chiral and large-$N_c$ limit, where one finds the simple expressions
[in the large-$N_c$ limit, $L_{22}^{W}$ scales as ${\cal O}(N_c)$ and becomes independent of the renormalization scale $\mu$,
and recall that ${\cal A}^3$ is also proportional to $N_c$, whereas $F_\pi$ scales as ${\cal O}(\sqrt{N_c})$]
\bea
\lim_{\genfrac{}{}{0pt}{}{m_q\to 0}{N_c\to \infty}} F_\pi {\widehat{\cal F}}_{\pi\gamma^*\gamma^*} (q_1^2,q_2^2) \!\!&=&\!\! {\cal A}^3 
- \frac{16}{3}   C_{22}^{W} (q_1^2 + q_2^2) + {\cal O}(p^6N_c^0 , p^8N_c)
\nonumber\\
\lim_{\genfrac{}{}{0pt}{}{m_q\to 0}{N_c\to \infty}} \! \left[ 
w_0^3 (q_1^2 , q_2^2 , (q_1+q_2)^2) + w_1^3 (q_1^2 , q_2^2 , (q_1+q_2)^2) 
\right] \!\!\!&=&\!\!\! 
\frac{{\cal A}^3 - (16/3) C_{22}^{W} (q_1^2 + q_2^2)}{(q_1 + q_2)^2} 
+ \frac{16}{3} C_{22}^{W} \! + {\cal O}(p^6N_c^0 , p^8N_c)   ,
\nonumber\\
\lim_{\genfrac{}{}{0pt}{}{m_q\to 0}{N_c\to \infty}} w_L^3 (q_1^2 , q_2^2 , (q_1+q_2)^2)
\!\!\!&=&\!\!\!
\frac{{\cal A}^3}{(q_1 + q_2)^2}   .
\lbl{comb_limit}
\eea
As is well known, there are no corrections to the above expression of $w_L^3 (q_1^2 , q_2^2 , (q_1+q_2)^2)$
in the chiral limit \cite{Adler:1969er,Witten:1983tw}. In the case of $w_L^3$, it is straightforward to understand how 
the limit in Eq. \rf{limit_q2_to_0} arises.
It simply reflects the fact that in the limit under consideration all that survives 
is the kinematic pole that has actually become a dynamical pion pole, with constant residue
fixed by the anomaly, and there is
nothing else, even before the limit $q_2\to 0$ is taken, as shown in the last expression 
in Eq. \rf{comb_limit}.
In the case of the sum $w_0^3+w_1^3$, the situation is somewhat more subtle.
There are other contributions besides a pion pole with constant residue in Eq. \rf{comb_limit} 
before the limit $q_2\to 0$ is taken: the momentum-dependent residue of the pole is given by whatever is left 
over from $F_\pi {\cal F}_{\pi\gamma^*\gamma^*} (q_1^2,q_2^2)$ in the combined
chiral and large-$N_c$ limit, i.e. here a contribution proportional to
$C_{22}^{W}$, and there are other, non-pole, contributions, also proportional to
$C_{22}^{W}$. When the limit $q_2\to 0$ is taken, these two different contributions combine such as to leave only 
\textit{a part} of the full pion pole, the one with a constant residue $F_0 \stackrel{\rm o}{{\cal F}}_{\pi\gamma^*\gamma^*} (0,0) = {\cal A}^3$, behind. 
That this will happen that way to higher, and in fact, to all orders in the low-energy expansion, is 
guaranteed by Eq. \rf{cond_w_i}, so that Eq. \rf{limit_q2_to_0} actually constitutes an exact result of QCD.
But as far as $w_0^3+w_1^3$ is concerned, it only reproduces a truncated part of the full pion pole
that was present to start with. In a nutshell, sometimes the two operations of taking the limit $q_2\to 0$ and of 
extracting the pion pole do not commute.

\end{document}